\documentclass[aps,prd,pdflatex,reprint,10pt,superscriptaddress,floatfix,nofootinbib]{revtex4-2}
\usepackage{float}
\usepackage{amsmath}
\usepackage[T1]{fontenc}
\usepackage[utf8]{inputenc}
\usepackage{lmodern}
\usepackage{amssymb}
\usepackage{natbib}
\usepackage{graphicx}
\usepackage{hyperref}
\usepackage{epstopdf}
\usepackage{pifont}
\usepackage{textgreek}
\usepackage[normalem]{ulem}
\usepackage{placeins}
\usepackage{tikz}
\usepackage{physics}
\usepackage{orcidlink}

\usepackage{booktabs,multirow,siunitx}
\usetikzlibrary{calc}
\graphicspath{{pdf/}}
\hypersetup{colorlinks=true,linkcolor=blue,citecolor=blue,filecolor=blue,urlcolor=blue,breaklinks=true}
\RequirePackage{color}

\begin{document}

\date{\today}

\title{Geometry-Controlled Exceptional Points in Helicoidal Orbital-Angular-Momentum Photonics}

\author{Edilberto O. Silva\orcidlink{0000-0002-0297-5747}}
\email[Edilberto O. Silva - ]{edilberto.silva@ufma.br}
\affiliation{Programa de P\'os-Gradua\c c\~ao em F\'{\i}sica \& Coordena\c c\~ao do Curso de F\'{\i}sica -- Bacharelado, Universidade Federal do Maranh\~{a}o, 65085-580 S\~{a}o Lu\'{\i}s, Maranh\~{a}o, Brazil}

\date{\today}

\begin{abstract}
Exceptional points (EPs) are non-Hermitian spectral singularities where eigenvalues and eigenvectors coalesce. We propose a geometry-controlled route to EPs in orbital-angular-momentum (OAM) photonics. In an effective torsion-free helicoidal background, the twist supplies a signed, real chiral splitting between opposite-OAM modes, while openness, gain, loss, leakage, or channel coupling, supplies the non-Hermiticity. The EP is reached when the geometric detuning compensates for a residual mode splitting and the linewidth contrast meets the non-Hermitian balance condition; the twist rate is thus the control parameter, not the source of non-Hermiticity. We obtain the OAM slope both for a confined quantum problem and, in the paraxial regime, for a helically twisted guide, and embed the OAM doublet in a multimode mode-space model to show that the EP survives spectator modes. Eigenvalue coalescence, phase-rigidity collapse, linear EP tracking with residual detuning, and branch exchange under geometric encircling provide falsifiable signatures. An OAM-resolved input-output response maps these onto spectroscopy-level observables, including spectral maps and three-dimensional Riemann-surface visualizations. A device-level paraxial simulation of two distinct photonic platforms, one reciprocal and one asymmetric-coupling Wiersig-type platform, reproduces the predicted exceptional point and its linear geometric tracking.
\end{abstract}

\maketitle

\section{Introduction}

Exceptional points (EPs) are defective degeneracies of non-Hermitian operators.
At an EP, two or more complex eigenvalues and their corresponding eigenvectors coalesce, producing branch-point spectra, non-orthogonal modal response, chiral state conversion, and enhanced sensitivity to perturbations~\cite{Berry2004,Heiss2012,Dembowski2001,Miri2019,Ozdemir2019,Bergholtz2021,Ding2022,Meng2024,Wiersig2014,Hodaei2017,Chen2017,Wiersig2020}.
They have been explored in optical cavities, waveguides, exciton-polariton platforms, acoustic systems, microwave resonators, and open wave systems.
Recent photonic work has also connected EPs directly to orbital-angular-momentum (OAM) conversion, synthetic OAM dimensions, and OAM microlasers~\cite{Allen1992,Andrews2012,Hayenga2019,Yang2023,Qi2024,Forbes2024}.

Two recent developments make the OAM setting especially relevant for the present proposal.
First, synthetic dimensions in photonics enable discrete mode labels, including frequency and OAM indices, to serve as controllable lattice coordinates beyond the ordinary real-space dimensionality~\cite{Ozawa2016,Yuan2018,Yang2023}.
Second, encircling an EP has become a direct means of revealing its Riemann-sheet topology, thereby enabling asymmetric mode switching, topological energy transfer, and OAM charge conversion~\cite{Doppler2016,Xu2016,Qi2024}.
The helicoidal splitting derived below naturally combines these two ingredients: it acts as a geometry-controlled potential along the OAM coordinate and therefore provides a physical parameter for moving and encircling EPs in an OAM synthetic space.

Most EP protocols tune non-Hermitian ingredients themselves: gain/loss contrast, dissipative coupling, radiative leakage, non-reciprocal coupling, or an externally imposed synthetic field~\cite{Bender1998,ElGanainy2018,Miri2019,Ozdemir2019}.
Here, we ask a narrower but useful question: can geometry tune the real detuning of an already-open system into the EP condition?
We answer this question affirmatively for the helicoidal background introduced in Sec.~\ref{secII}. The non-Hermitian part is still supplied by openness, gain, loss, or coupling to external channels. The new ingredient is that the helicoidal twist supplies a controllable real chiral splitting between opposite-OAM modes.
This distinction is important: the proposal is not that geometry alone creates a non-Hermitian degeneracy, but that geometry becomes the parameter that steers an open two-mode system to the defective point.

In a closed quantum system, the helicoidal twist produces two separable effects.
First, it induces harmonic radial confinement controlled by the positive scale $|\Omega k_z|$.
Second, it produces a signed Zeeman-like shift proportional to $-m\Omega k_z$, thereby lifting the $m\leftrightarrow -m$ degeneracy at fixed axial wave number.
In a realistic device, the selected doublet will generally also have a residual non-geometric detuning produced by static index detuning, boundary deformation, residual anisotropy, pump-profile asymmetry, OAM-selective scattering, or fabrication imperfections.
The operational mechanism is therefore compensation: the twist shifts the real detuning until the open two-mode system reaches the EP condition.
Thus, the experimentally relevant claim is not merely that twist splits OAM modes, but that the measured EP location should follow the geometric slope derived below.
This gives direct tests: changing the residual detuning should move the EP linearly, reversing the handedness should reverse the OAM ordering, and changing $m$, $n_z$, or $L$ should rescale the compensation point.

\begin{figure*}[t]
\centering
\includegraphics[scale=0.4]{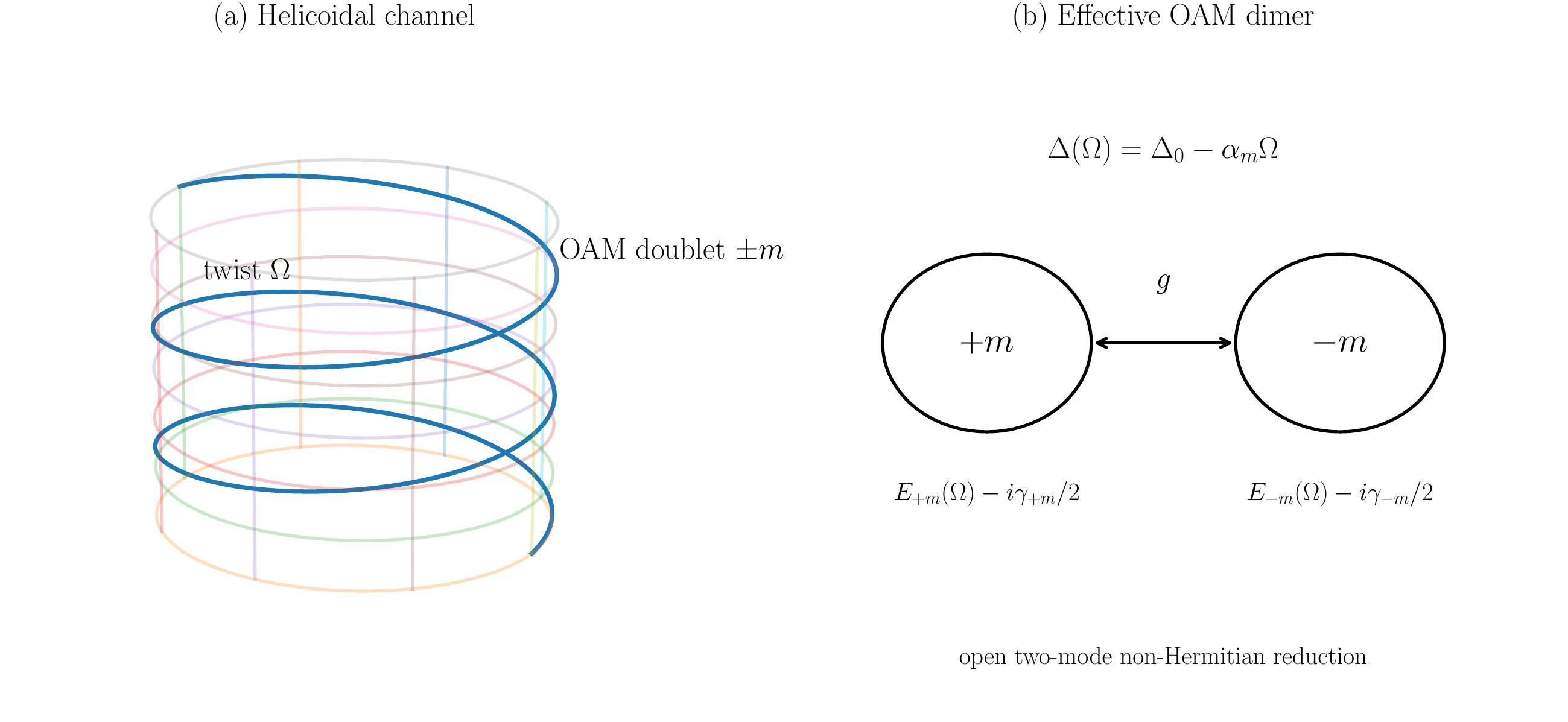}
\caption{Conceptual structure of a geometry-controlled exceptional point.
(a) A helicoidal channel or effective helicoidal medium couples axial and azimuthal motion through the twist parameter $\Omega$.
(b) Projection onto two opposite-OAM modes gives a non-Hermitian two-mode problem with coupling $g$, linewidths $\gamma_{+m}$ and $\gamma_{-m}$, and real detuning $\Delta(\Omega)=\Delta_0-\alpha_m\Omega$.}
\label{fig:schematic}
\end{figure*}

The paper is organized as follows.
In Sec.~\ref{secII}, we review the helicoidal metric, define the regime of validity, and obtain the corresponding chiral spectrum.
In Sec.~\ref{secIII}, we construct the two-mode non-Hermitian reduction.
In Sec.~\ref{secIV}, we derive the exceptional-point condition and identify the twist-controlled detuning.
Section~\ref{secV} gives the minimal two-mode numerical illustration.
Section~\ref{secVI} presents the OAM-mode-space simulations, including multimode robustness, phase rigidity, geometric EP tracking, and both static and dynamical encircling.
Section~\ref{secVII} presents a step-by-step photonic OAM implementation, along with the calibration protocol and falsifiable signatures.
Section~\ref{secVIII} validates the mechanism with a device-level paraxial simulation of two distinct photonic platforms. Section~\ref{secIX} connects the theory to experimentally accessible cavity measurements through an OAM-resolved input-output response, three-dimensional spectral visualizations, and the photonic observables and accessible scales.
Section~\ref{secX} discusses scope, limitations, and outlook, and Sec.~\ref{secXI} concludes.

\section{Helicoidal metric, validity, and chiral spectrum}\label{secII}

We consider the helicoidal metric
\begin{equation}
ds^2=dr^2+r^2d\phi^2+\left(dz+\Omega r^2d\phi\right)^2.
\label{metric2}
\end{equation}
The parameter $\Omega$ has dimensions of inverse length, so that $\Omega r^2d\phi$ has the same dimension as $dz$.
It is a metric twist parameter, not Cartan torsion.
All geometric quantities in this work are built from the Levi-Civita connection, and the Riemann-Cartan torsion tensor is therefore identically zero.
The term ``helicoidal twist'' will be used in this metric sense throughout. 
The background is nonetheless genuinely curved rather than a pure coordinate artifact: the Levi-Civita Riemann tensor has nonvanishing components and the Ricci scalar is constant and negative, $R=-2\Omega^2$. This distinguishes the helicoidal background from a flat space dressed by a synthetic gauge field. At the same time, the chiral splitting derived below originates specifically from the off-diagonal metric component $g^{\phi z}=-\Omega$, i.e.\ from minimal coupling to the synthetic potential one-form $A=\Omega r^2 d\phi$, and is therefore an Aharonov-Bohm/Zeeman-like term rather than a direct curvature effect. We keep this distinction explicit: curvature characterizes the background, whereas the gauge-like cross term controls the OAM detuning.

The model should be understood as an effective single-particle or scalar-wave description in a helicoidal background, in the same spirit as effective geometric descriptions used for quantum motion in curved nanostructures and constrained systems~\cite{DaCosta1981,Ortix2015,Gentile2022,Silva2026}.
It assumes that the relevant dynamics can be projected onto separable axial and azimuthal channels, that the selected two modes are spectrally isolated from other modes, and that openness can be represented by finite linewidths and effective intermodal couplings. For definiteness, we take $k_z>0$ and keep two frequency scales explicit below: a positive confinement frequency set by $|\Omega k_z|$ and a signed chiral frequency proportional to $\Omega k_z$.
This convention keeps the radial oscillator well defined for either handedness of the helicoidal twist while preserving the sign of the OAM splitting.

In coordinates $(r,\phi,z)$,
\begin{equation}
g_{ij}=
\begin{pmatrix}
1 & 0 & 0 \\
0 & r^2+\Omega^2 r^4 & \Omega r^2 \\
0 & \Omega r^2 & 1
\end{pmatrix},
\qquad
\sqrt{g}=r.
\end{equation}
The inverse metric is
\begin{equation}
\begin{gathered}
g^{rr}=1,\qquad
g^{\phi\phi}=\frac{1}{r^2},\qquad
g^{\phi z}=g^{z\phi}=-\Omega,\\
g^{zz}=1+\Omega^2r^2.
\end{gathered}
\end{equation}
The corresponding Laplace-Beltrami operator is
\begin{equation}
\nabla^2_{\rm LB}
=
\frac{1}{r}\partial_r\left(r\partial_r\right)
+
\frac{1}{r^2}\partial_\phi^2
+
\left(1+\Omega^2r^2\right)\partial_z^2
-
2\Omega \partial_\phi\partial_z .
\label{LB}
\end{equation}
For a spinless particle of effective mass $m^\ast$, the Schr\"odinger equation is
\begin{equation}
-\frac{\hbar^2}{2m^\ast}\nabla^2_{\rm LB}\Psi=E\Psi.
\end{equation}
We use the separable ansatz
\begin{equation}
\Psi(r,\phi,z)=e^{im\phi}e^{ik_z z}F(r),
\qquad
m\in\mathbb Z,
\label{ansatz}
\end{equation}
with axial quantization
\begin{equation}
k_z=\frac{\pi n_z}{L},
\qquad
n_z=1,2,\ldots .
\end{equation}
Substitution into Eq.~\eqref{LB} gives the radial equation
\begin{align}
-\frac{\hbar^2}{2m^\ast}
\left(
F''+\frac{1}{r}F'
\right)
+
\frac{\hbar^2m^2}{2m^\ast r^2}F
+
\frac{\hbar^2\Omega^2k_z^2}{2m^\ast}r^2F
\nonumber\\
-
\frac{\hbar^2\Omega k_z m}{m^\ast}F
+
\frac{\hbar^2k_z^2}{2m^\ast}F
=
EF.
\label{radial_raw}
\end{align}
After the standard transformation $F(r)=u(r)/\sqrt r$, one obtains the effective one-dimensional radial problem
\begin{equation}
-\frac{\hbar^2}{2m^\ast}\frac{d^2u}{dr^2}
+
V_{\rm eff}(r)u
=
Eu,
\end{equation}
where
\begin{equation}
V_{\rm eff}(r)
=
\frac{\hbar^2}{2m^\ast}\frac{m^2-\frac14}{r^2}
+
\frac12m^\ast\omega_c^2 r^2
-
\hbar\omega_s m
+
\frac{\hbar^2k_z^2}{2m^\ast}.
\label{Veff}
\end{equation}
Here
\begin{equation}
\omega_c=\frac{\hbar |\Omega k_z|}{m^\ast},
\qquad
\omega_s=\frac{\hbar \Omega k_z}{m^\ast}
=
\frac{\hbar\Omega\pi n_z}{m^\ast L}
\label{omega_scales}
\end{equation}
are, respectively, the positive radial confinement frequency and the signed chiral frequency. The first term in Eq.~\eqref{Veff} is the Langer-corrected centrifugal barrier, the second is the geometry-induced harmonic confinement, the third is the chiral Zeeman-like shift, and the fourth is the axial kinetic offset.

The normalized one-dimensional radial functions are
\begin{align}
u_{n_r,m}(r)&=
{\cal N}_{n_r,m}\,
r^{|m|+1/2}
e^{-\xi/2}
L_{n_r}^{|m|}(\xi),\nonumber\\
\xi&=\frac{m^\ast\omega_c}{\hbar}r^2 .
\end{align}
The original radial envelope in Eq.~\eqref{ansatz} is $F_{n_r,m}(r)=u_{n_r,m}(r)/\sqrt r$.
Thus, the equivalent normalization conventions are
\begin{equation}
\int_0^\infty |u_{n_r,m}(r)|^2\,dr=1,
\qquad
\int_0^\infty r|F_{n_r,m}(r)|^2\,dr=1 .
\end{equation}
The normalization constant is
\begin{equation}
{\cal N}_{n_r,m}
=
\left[
\frac{
2\left(m^\ast\omega_c/\hbar\right)^{|m|+1}n_r!
}{
\Gamma(n_r+|m|+1)
}
\right]^{1/2}.
\label{Norm}
\end{equation}
The corresponding energy spectrum is
\begin{equation}
E_{n_r,m,n_z}
=
\hbar\omega_{c,n_z}\left(2n_r+|m|+1\right)
+
\frac{\hbar^2\pi^2n_z^2}{2m^\ast L^2}
-
\hbar\omega_{s,n_z}m,
\label{Spectrum}
\end{equation}
with
\begin{equation}
\omega_{c,n_z}=\frac{\hbar |\Omega|\pi n_z}{m^\ast L},
\qquad
\omega_{s,n_z}=\frac{\hbar\Omega\pi n_z}{m^\ast L}.
\end{equation}
Equation~\eqref{Spectrum} is the starting point of the present work.
It shows that helicoidal geometry produces a real, tunable, chiral detuning between states with different azimuthal quantum numbers.

For two states with the same $n_r$ and $n_z$, but opposite angular momenta $m$ and $-m$, the energy difference is
\begin{align}
\Delta_{\rm geom}(\Omega)
&=
E_{n_r,m,n_z}-E_{n_r,-m,n_z} \nonumber\\
&=
-2m\hbar\omega_{s,n_z}
=
-\frac{2m\hbar^2\Omega\pi n_z}{m^\ast L}.
\label{DeltaGeom}
\end{align}
This is the geometric detuning used to control the exceptional point.

\subsection{Consistency checks and two-mode isolation}

Several points are worth making explicit before introducing non-Hermiticity.
First, Eq.~\eqref{LB} follows directly from $\nabla^2_{\rm LB}=g^{-1/2}\partial_i(g^{1/2}g^{ij}\partial_j)$ with $\sqrt g=r$; no torsional contribution is present because the connection is Levi-Civita.
Second, the separability used in Eq.~\eqref{ansatz} is exact for the metric in Eq.~\eqref{metric2}, since $\phi$ and $z$ are cyclic coordinates.
Third, the oscillator scale in the radial wave function is manifestly positive because it is controlled by $\omega_{c,n_z}$, whereas the OAM splitting is controlled by the signed quantity $\omega_{s,n_z}$.
This separation of scales makes the formulas invariant under reversal of the helicoidal handedness: changing $\Omega\to-\Omega$ leaves the radial confinement unchanged but reverses the ordering of the $m$ and $-m$ states.
Finally, the two-mode reduction below requires the chosen pair to be well separated from all other helicoidal modes, namely
\begin{equation}
|E_j-E_{1,2}|\gg |g_{12}|,|g_{21}|,\gamma_1,\gamma_2
\end{equation}
for every neglected mode $j$.
If this condition is relaxed, the same geometric-detuning mechanism persists, but the exceptional point may be embedded in a larger non-Hermitian mode network rather than described by a single isolated square-root branch point.

\section{Two-mode non-Hermitian reduction}\label{secIII}

We now open the helicoidal system by allowing the relevant modes to have finite linewidths and to be mutually coupled. The physical origin of non-Hermiticity depends on the platform.
In photonics, it may arise from absorption, gain, radiation leakage, or deliberately engineered loss channels. More formally, projecting an open system onto an internal subspace gives an energy-dependent effective Hamiltonian of Feshbach type, $H_{\rm eff}(E)=H_B+V(E^+-H_C)^{-1}V^\dagger$, whose wide-band or weakly energy-dependent limit has the familiar form $H_B-iWW^\dagger/2$~\cite{Rotter2009,RotterBird2015}.
In either case, near an isolated doublet, the minimal description is a two-mode effective Hamiltonian.

Let $|1\rangle$ and $|2\rangle$ be two helicoidal modes.
For concreteness, one may take
\begin{equation}
|1\rangle=|n_r,m,n_z\rangle,
\qquad
|2\rangle=|n_r,-m,n_z\rangle,
\end{equation}
although the formalism also applies to nearby modes with different $n_r$, $m$, or $n_z$.
The effective Hamiltonian is
\begin{equation}
H_{\rm eff}
=
\begin{pmatrix}
E_1(\Omega)-i\gamma_1/2 & g_{12} \\
g_{21} & E_2(\Omega)-i\gamma_2/2
\end{pmatrix}.
\label{Heff}
\end{equation}
Here $E_{1,2}(\Omega)$ are the real helicoidal energies, $\gamma_{1,2}$ are linewidths, and $g_{12},g_{21}$ are effective intermodal couplings.

It is useful to introduce
\begin{equation}
E_0=\frac{E_1+E_2}{2},
\qquad
\Delta(\Omega)=E_1(\Omega)-E_2(\Omega),
\end{equation}
and
\begin{equation}
\bar{\gamma}=\frac{\gamma_1+\gamma_2}{2},
\qquad
\Delta\gamma=\gamma_1-\gamma_2.
\end{equation}
Then Eq.~\eqref{Heff} can be written as
\begin{equation}
H_{\rm eff}
=
\left(E_0-\frac{i}{2}\bar{\gamma}\right)\mathbb I
+
\begin{pmatrix}
\frac{\Delta}{2}-\frac{i}{4}\Delta\gamma & g_{12}\\
g_{21} & -\frac{\Delta}{2}+\frac{i}{4}\Delta\gamma
\end{pmatrix}.
\label{HeffShifted}
\end{equation}
The complex eigenvalues are
\begin{equation}
\lambda_\pm
=
E_0-\frac{i}{2}\bar{\gamma}
\pm
\frac12
\sqrt{
\left(\Delta-\frac{i}{2}\Delta\gamma\right)^2
+
4g_{12}g_{21}
}.
\label{Eigenvalues}
\end{equation}
An exceptional point occurs when the square root vanishes:
\begin{equation}
\left(\Delta(\Omega)-\frac{i}{2}\Delta\gamma\right)^2
+
4g_{12}g_{21}
=0.
\label{EP_general}
\end{equation}
This is the central condition of the paper.

In the simplest reciprocal case, $g_{12}=g_{21}=g$ with real $g$, Eq.~\eqref{EP_general} implies
\begin{equation}
\Delta(\Omega_{\rm EP})=0,
\qquad
|\Delta\gamma|=4|g|.
\label{EP_recip}
\end{equation}
Thus, for reciprocal coupling, non-Hermiticity fixes the required linewidth contrast, while geometry tunes the real detuning to zero.
For an ideal opposite-OAM pair with no additional detuning, this zero occurs at the symmetry point of the detuning.
In an actual platform, however, static confinement, boundary deformation, strain, gates, refractive-index offsets, or contact asymmetry generically produce a residual non-geometric detuning $\Delta_0$.
The relevant detuning should then be written as
\begin{equation}
\Delta(\Omega)=\Delta_0+\Delta_{\rm geom}(\Omega).
\label{DeltaTotal}
\end{equation}
For the opposite-OAM pair, Eq.~\eqref{DeltaGeom} gives
\begin{equation}
\Delta(\Omega)
=
\Delta_0
-
\frac{2m\hbar^2\Omega\pi n_z}{m^\ast L}.
\label{DeltaTotal2}
\end{equation}
The corresponding finite-twist exceptional point is obtained by imposing this compensation together with the linewidth-balance condition in Eq.~\eqref{EP_recip}.
The twist does not need to generate the non-Hermiticity; it supplies the real compensation that places the open system at the defective degeneracy.

More general exceptional points occur when $g_{12}g_{21}$ is complex.
This situation can arise in non-reciprocal mode coupling, dissipative coupling, or gain-assisted photonic structures.
Writing
\begin{equation}
g_{12}g_{21}=|G|e^{i\theta_G},
\end{equation}
the exceptional-point condition becomes
\begin{equation}
\left(\Delta-\frac{i}{2}\Delta\gamma\right)^2
=
-4|G|e^{i\theta_G}.
\end{equation}
In this case, both the real and imaginary parts of the degeneracy condition can be satisfied at a nonzero geometric detuning even without requiring $\Delta(\Omega)=0$.
Thus, non-reciprocal or dissipative couplings enlarge the parameter space in which helicoidal twist can drive the system through an exceptional point.

\section{Exceptional points from helicoidal twist}\label{secIV}

We now specialize in the geometrically split opposite-OAM pair.
The relevant detuning is
\begin{equation}
\Delta_{\rm geom}(\Omega)
=
-\alpha_m \Omega,
\qquad
\alpha_m=
\frac{2m\hbar^2\pi n_z}{m^\ast L},
\label{alpha}
\end{equation}
where $m>0$ labels the magnitude of the two opposite angular momenta.
Including a platform-dependent detuning $\Delta_0$, the total real detuning is
\begin{equation}
\Delta(\Omega)=\Delta_0-\alpha_m\Omega.
\label{DeltaComp}
\end{equation}

\subsection{Geometric compensation of residual detuning}

Equation~\eqref{DeltaComp} makes explicit the operational meaning of geometry-controlled exceptional-point tuning. The helicoidal twist is not required to be the origin of gain, loss, or linewidth imbalance. Those ingredients enter through $\Delta\gamma$ and the non-Hermitian structure of $H_{\rm eff}$. Instead, $\Omega$ controls the real part of the two-mode detuning. A finite residual detuning $\Delta_0$ may be produced deliberately or unintentionally by boundary ellipticity, refractive-index contrast, pump-profile asymmetry, OAM-selective scattering, or imperfect confinement. The exceptional point is reached when the geometric contribution compensates for this residual splitting,
\begin{equation}
\Delta_0-\alpha_m\Omega_{\rm EP}=0.
\label{CompensationCondition}
\end{equation}
Thus, the finite-twist exceptional point is
\begin{equation}
\Omega_{\rm EP}=\frac{\Delta_0}{\alpha_m}
=
\frac{m^\ast L}{2m\hbar^2\pi n_z}\Delta_0,
\label{OmegaEPSection}
\end{equation}
provided the non-Hermitian condition on linewidths and coupling is also satisfied.
This formulation avoids a possible ambiguity of the perfectly symmetric case: if $\Delta_0=0$ and the coupling is reciprocal and real, the real-detuning condition is satisfied only at the symmetry point of the detuning, whereas a realistic device with $\Delta_0\neq0$ can be brought to the exceptional point by tuning a nonzero helicoidal twist.

The eigenvalue splitting is therefore
\begin{equation}
\lambda_+-\lambda_-
=
\sqrt{
\left(
\Delta_0-\alpha_m\Omega-\frac{i}{2}\Delta\gamma
\right)^2
+
4g^2
},
\label{splitting}
\end{equation}
for the reciprocal case.

Equation~\eqref{splitting} shows the square-root topology of the exceptional point.
Near $\Omega_{\rm EP}$, define
\begin{equation}
\delta\Omega=\Omega-\Omega_{\rm EP}.
\end{equation}
At the loss-balance condition $|\Delta\gamma|=4|g|$, the splitting scales as
\begin{equation}
\lambda_+-\lambda_-
\propto
\sqrt{\delta\Omega}.
\label{sqrt}
\end{equation}
This non-analytic dependence is characteristic of the exceptional-point response.
The key distinction of the present proposal is that the perturbation parameter is geometric: it is the helicoidal twist rate $\Omega$.

The eigenvectors also coalesce at the exceptional point.
For the shifted Hamiltonian in Eq.~\eqref{HeffShifted}, an eigenvector associated with $\lambda_\pm$ may be written as
\begin{equation}
|\psi_\pm\rangle
=
\begin{pmatrix}
1\\
\dfrac{
-\left(\Delta-\frac{i}{2}\Delta\gamma\right)
\pm
\sqrt{
\left(\Delta-\frac{i}{2}\Delta\gamma\right)^2+4g^2
}
}{2g}
\end{pmatrix}.
\end{equation}
At the exceptional point, the square root vanishes, and the two eigenvectors become parallel:
\begin{equation}
|\psi_+\rangle \parallel |\psi_-\rangle .
\end{equation}
Thus, the helicoidal twist controls not merely a level crossing, but a genuine non-Hermitian coalescence.

\subsection{Dimensionless form used in the figures}

For the reciprocal model, it is convenient to remove the irrelevant common shift $E_0-i\bar\gamma/2$ and measure all splittings in units of $g$.
Defining
\begin{equation}
\delta=\frac{\Delta}{g},
\qquad
\eta=\frac{\Delta\gamma}{4g},
\qquad
x=\frac{\Omega}{\Omega_{\rm EP}},
\end{equation}
the shifted eigenvalues are
\begin{equation}
\tilde\lambda_\pm
=
\frac{\lambda_\pm-E_0+i\bar\gamma/2}{g}
=
\pm\frac12\sqrt{\left(\delta-2i\eta\right)^2+4}.
\label{dimensionless_lambdas}
\end{equation}
In the compensation geometry, $\delta=(\Delta_0/g)(1-x)$.
The exceptional point in Fig.~\ref{fig:eigenvalues} corresponds to $x=1$ and $\eta=1$.
Equation~\eqref{dimensionless_lambdas} also makes clear why a single mechanical or synthetic variation of $\Omega$ is insufficient by itself to reach a generic EP: the real detuning is supplied by geometry, while the imaginary detuning must be set by loss, gain, leakage, or coupling to external channels.

\section{Numerical illustration of the geometry-controlled exceptional point}\label{secV}

\begin{figure*}[t]
\centering
\includegraphics[scale=0.37]{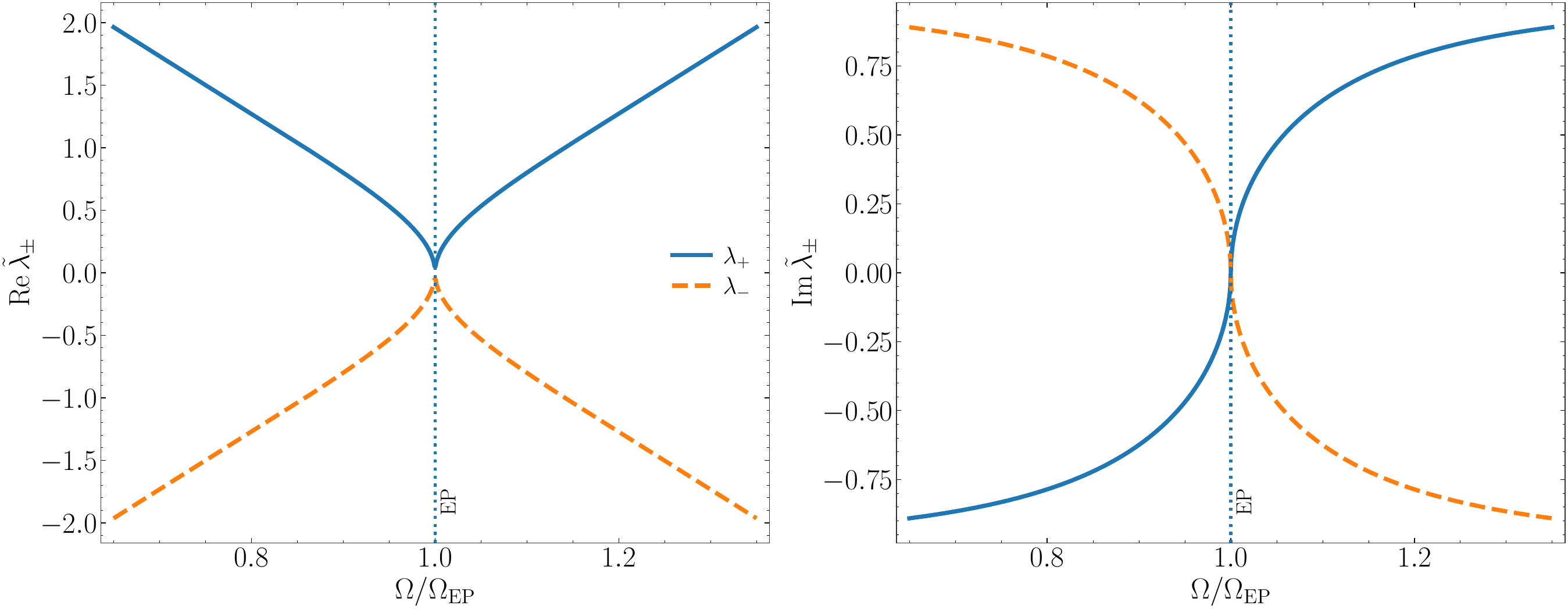}
\caption{Dimensionless eigenvalue branches near a geometry-controlled exceptional point. The plotted quantity is $\tilde\lambda_\pm=(\lambda_\pm-E_0+i\bar\gamma/2)/g$ for $g=0.10~{\rm meV}$, $\Delta\gamma=4g$, and $\Delta(\Omega)=\Delta_0(1-\Omega/\Omega_{\rm EP})$. The two eigenvalues and eigenvectors coalesce at $\Omega=\Omega_{\rm EP}$.}
\label{fig:eigenvalues}
\end{figure*}

We now give a minimal numerical illustration of the preceding analytical result.
The purpose is not to simulate a specific device geometry, but to display the universal square-root coalescence that occurs when the twist tunes the real detuning via the exceptional-point condition.
We use the dimensionless form of Eq.~\eqref{dimensionless_lambdas} and take
\begin{equation}
g=0.10~{\rm meV},
\qquad
\Delta\gamma=4g,
\qquad
\Delta_0=1.00~{\rm meV},
\end{equation}
and use the normalized twist coordinate
\begin{equation}
x=\frac{\Omega}{\Omega_{\rm EP}},
\qquad
\Delta(\Omega)=\Delta_0(1-x).
\end{equation}
In this parametrization, the exceptional point occurs at $x=1$.
Figure~\ref{fig:eigenvalues} shows that both branches of the complex spectrum coalesce at this point.
The real and imaginary parts exchange their square-root character across the degeneracy, as expected from Eq.~\eqref{splitting}.

The energy unit (meV) is used here only to make the dimensionless ratios concrete; since all splittings are measured in units of $g$, the example is platform-independent. In a photonic realization, the same numbers map onto frequency units: e.g.\ taking $g/2\pi=2~{\rm GHz}$ gives $\Delta\gamma/2\pi=4g/2\pi=8~{\rm GHz}$ and a residual detuning $\Delta_0/2\pi=20~{\rm GHz}$, with the twist coordinate calibrated from the measured OAM splitting slope.
In a device, these parameters are extracted from measured resonance frequencies, propagation constants, and linewidths rather than being imposed.
The essential prediction is therefore not the absolute value of $\Omega_{\rm EP}$ in a particular material, but the compensation law $\Omega_{\rm EP}=\Delta_0/\alpha_m$ and its linear displacement when $\Delta_0$ is varied.

\begin{figure}[t]
\centering
\includegraphics[scale=0.34]{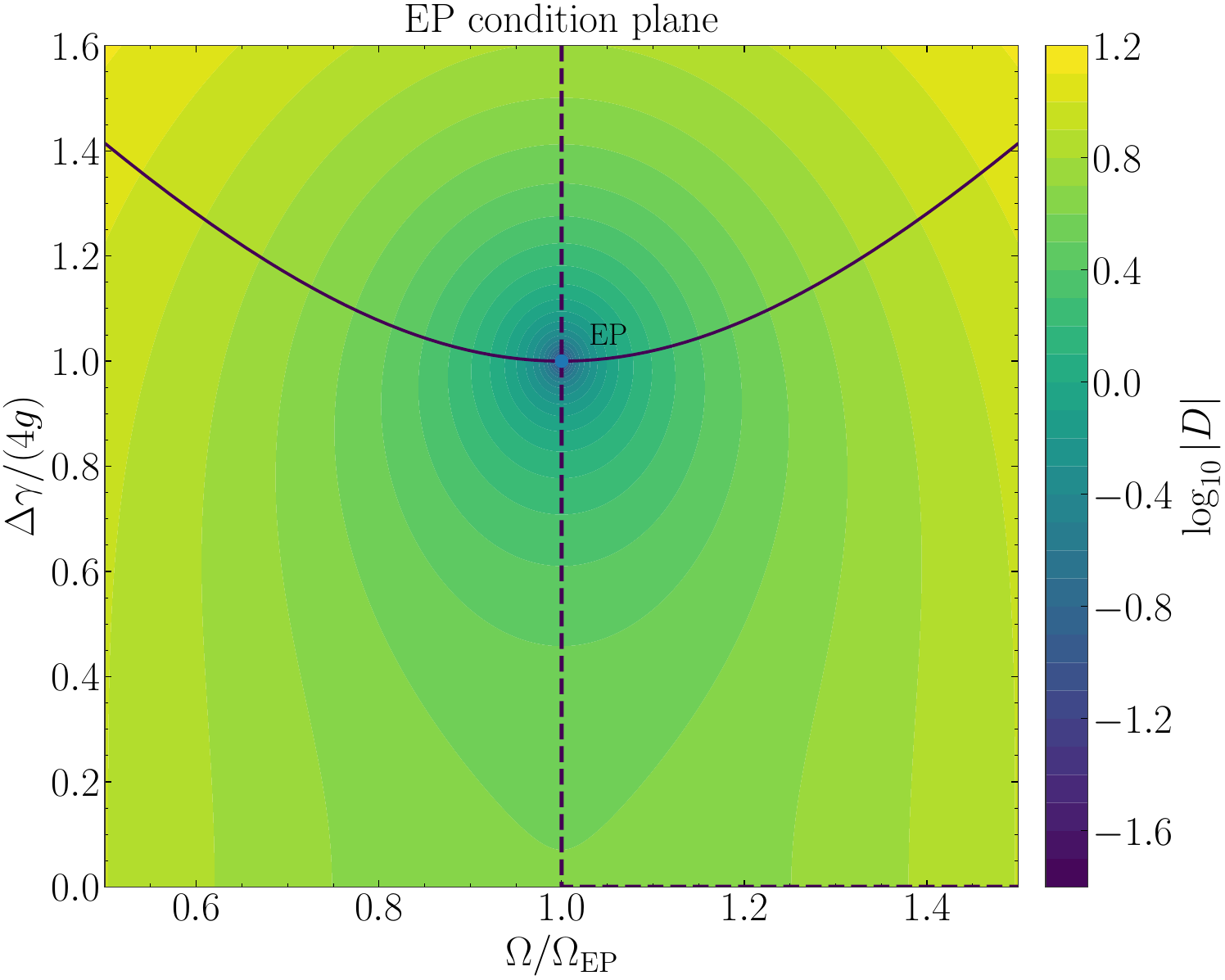}
\caption{Exceptional-point condition in the reciprocal model. The discriminant $D=(\Delta-i\Delta\gamma/2)^2+4g^2$ vanishes only when the real geometric detuning is compensated, $\Delta=0$, and the linewidth contrast satisfies $|\Delta\gamma|=4|g|$. The solid line is ${\rm Re}\,D=0$, the dashed line is ${\rm Im}\,D=0$, and the marked intersection gives the exceptional point.}
\label{fig:epplane}
\end{figure}

The reciprocal real-coupling exceptional-point condition can also be visualized in the two-parameter plane spanned by normalized twist and loss contrast.
Let
\begin{equation}
D(\Omega,\Delta\gamma)=
\left(
\Delta_0-\alpha_m\Omega-\frac{i}{2}\Delta\gamma
\right)^2+4g^2
\end{equation}
be the discriminant under the square root.
The exceptional point is the simultaneous zero of ${\rm Re}\,D$ and ${\rm Im}\,D$.
Figure~\ref{fig:epplane} shows this intersection in the $(\Omega/\Omega_{\rm EP},\Delta\gamma/4g)$ plane.
The solid contour denotes ${\rm Re}\,D=0$, the dashed contour denotes ${\rm Im}\,D=0$, and their intersection marks the exceptional point.
This representation makes clear that geometry alone tunes the horizontal coordinate, whereas openness and loss imbalance tune the vertical coordinate.

\section{OAM-mode-space simulations and geometric EP topology}\label{secVI}

The two-mode Hamiltonian analytically isolates the defective degeneracy, but a photonic proposal should also demonstrate that the mechanism persists within a larger OAM manifold. We therefore recast the helicoidal splitting as a field in OAM mode space and then test the target exceptional point against spectator modes, phase rigidity, static branch exchange, and dynamical encircling. The word ``synthetic'' is used here in the standard modal sense: the OAM modes are physical optical modes, while the discrete index $\ell$ is treated as a lattice coordinate in mode space.

\subsection{OAM-mode-space representation}

In a degenerate cavity, ring resonator, or waveguide platform, the OAM index can be used as a discrete synthetic coordinate rather than merely as an internal label~\cite{Yang2023,Ozawa2016,Yuan2018}.
A generic OAM-space Hamiltonian is
\begin{align}
H_{\rm OAM}
&=
\sum_{\ell}
\left[
\varepsilon^{(0)}_{\ell}
-
\beta_{\rm geo}\Omega\ell
-
\frac{i}{2}\gamma_\ell
\right]
|\ell\rangle\langle\ell|
\notag\\&+
\sum_{\ell,q}
\left(
J_{\ell,q}|\ell+q\rangle\langle\ell|+{\rm H.c.}
\right).
\label{HOAM}
\end{align}
Here $\ell$ labels the OAM mode, $J_{\ell,q}$ is generated by an angular grating or mode converter carrying angular momentum $q$, and $\gamma_\ell$ is an OAM-dependent linewidth.
The coefficient $\beta_{\rm geo}$ is the measured photonic analog of the helicoidal slope.
For the quantum spectrum derived in Sec.~\ref{secII}, $\beta_{\rm geo}\Omega\ell$ corresponds to $\hbar\omega_s\ell$. Explicitly, the diagonal term $-\beta_{\rm geo}\Omega\ell$ matches the chiral shift $-\hbar\omega_s\ell$ provided $\beta_{\rm geo}=\hbar\omega_s/\Omega=\hbar^2 k_z/m^\ast$; its sign reverses under $\Omega\to-\Omega$, consistent with handedness reversal. In an optical realization, the same term should be read as a propagation-constant or resonance-frequency shift calibrated from the OAM splitting.

Keeping the pair $\ell=\pm m$ and a coupling channel with $q=2m$ reduces Eq.~\eqref{HOAM} to Eq.~\eqref{Heff}. The advantage of Eq.~\eqref{HOAM} is that it also makes clear what must be checked experimentally: the selected OAM doublet must remain isolated from nearby $\ell$ modes over the range of twist and loss used to locate the EP. The helicoidal term is then not an arbitrary detuning parameter; it is a synthetic-coordinate potential whose sign changes under handedness reversal and whose slope scales with the chosen OAM sector.

\subsection{Multimode spectral test}

\begin{figure*}[t]
\centering
\includegraphics[scale=0.39]{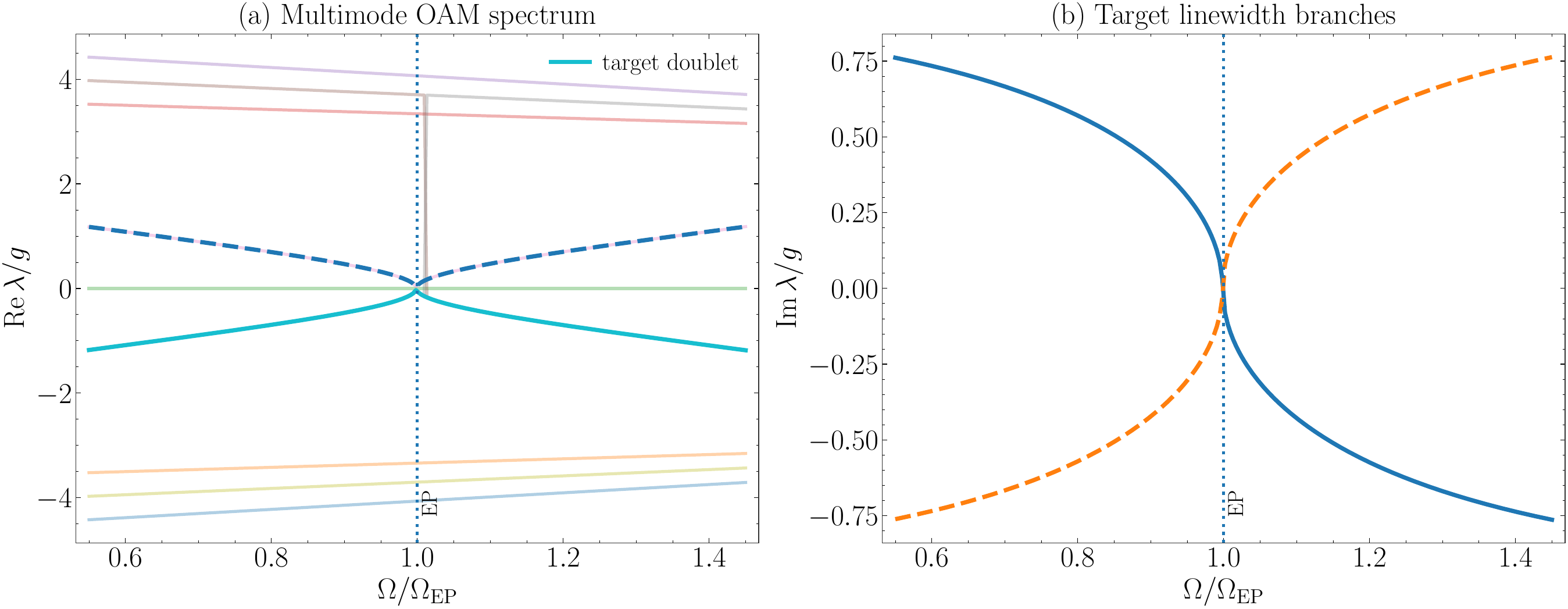}
\caption{Multimode OAM-space test of the geometry-controlled EP.
(a) Real parts of the eigenvalues of a finite OAM manifold with modes $\ell=-4,\ldots,4$. The thick branches are the target $\ell=\pm1$ doublet, while the thin branches are spectator OAM modes. The target branches coalesce at the calibrated twist even when weak residual couplings to spectator modes are present.
(b) Imaginary parts of the two target branches, showing the associated linewidth coalescence.}
\label{fig:multimode}
\end{figure*}

Figure~\ref{fig:multimode} shows a finite OAM-space calculation with spectator modes $\ell=-4,\ldots,4$. The target pair is $\ell=\pm1$, coupled by an angular perturbation with $q=2$. Additional weak $q=2$ couplings to neighboring spectator modes are included to test the stability of the two-mode reduction. The diagonal energies of the target pair are chosen so that their real detuning obeys $\Delta(\Omega)=\Delta_0(1-\Omega/\Omega_{\rm EP})$, while the linewidth contrast satisfies the reciprocal EP condition. The spectator modes are detuned by several units of $g$.

This calculation turns the two-mode assumption into a numerical diagnostic.
If the spectator modes remain separated, the target doublet retains the square-root EP, and its location follows the same compensation law. If a spectator mode approaches within the scale set by $g$ or $\gamma_\ell$, the observed singularity is no longer a clean second-order EP, and the system must be treated as a multimode non-Hermitian network. Thus, the multimode calculation is not a separate physical hypothesis; it is a validation step for the OAM-space implementation.

\subsection{Phase rigidity and geometric tracking}

\begin{figure*}[t]
\centering
\includegraphics[scale=0.39]{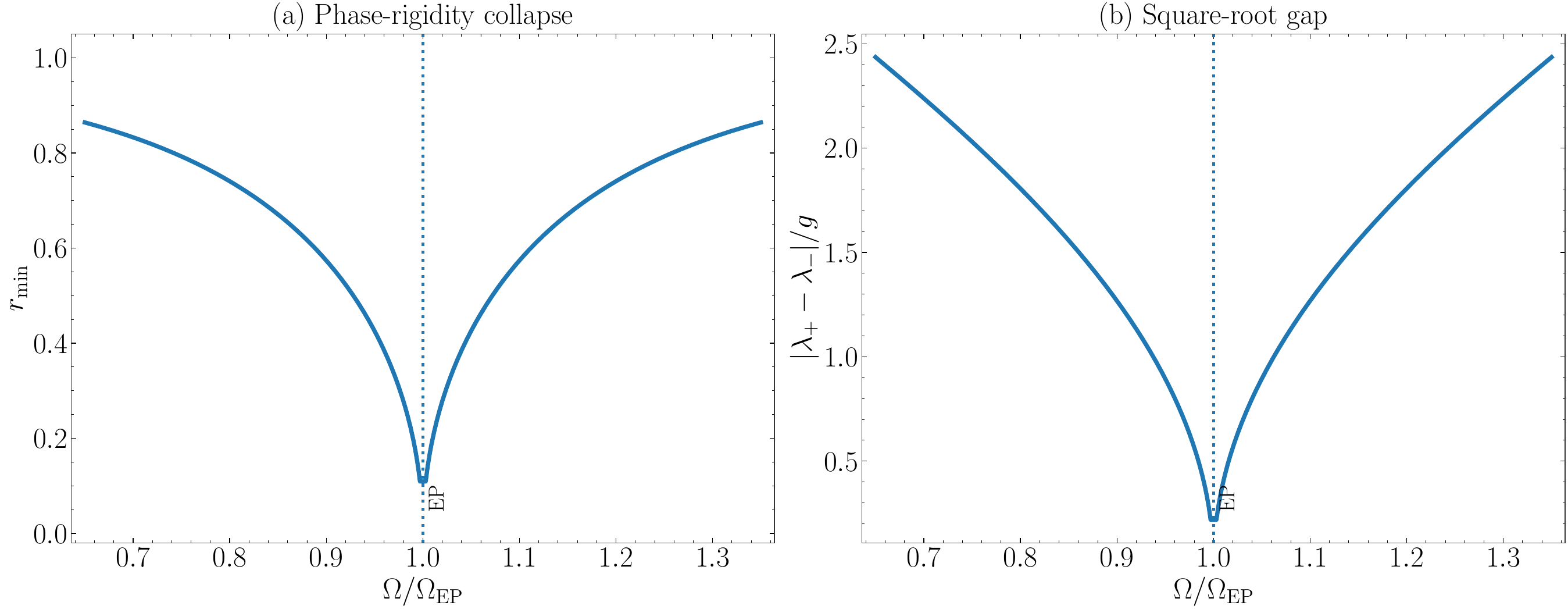}
\caption{Eigenvector-sensitive verification of the geometry-controlled EP.
(a) Minimum phase rigidity of the reciprocal complex-symmetric two-mode model as a function of normalized twist. The collapse near $\Omega=\Omega_{\rm EP}$ indicates eigenvector coalescence rather than a conventional crossing. (b) Corresponding eigenvalue gap, displaying the square root closing at the same twist.}
\label{fig:rigidity}
\end{figure*}

Eigenvalue coalescence alone is not sufficient to establish an exceptional point.
For a reciprocal complex-symmetric two-mode reduction, an eigenvector-sensitive diagnostic is the phase rigidity
\begin{equation}
r_j=
\frac{|\psi_j^{T}\psi_j|}{\psi_j^\dagger\psi_j},
\label{phase_rigidity_simple}
\end{equation}
where $\psi_j$ is a right eigenvector of the shifted non-Hermitian Hamiltonian.
At an ideal second-order EP, the two eigenvectors become self-orthogonal in the complex-symmetric bilinear product and $r_j$ vanishes. Equation~\eqref{phase_rigidity_simple} assumes a complex-symmetric reduction ($g_{12}=g_{21}$); for non-reciprocal coupling, the appropriate diagnostic is the left/right-eigenvector form of Eq.~\eqref{phase_rigidity}.
Figure~\ref{fig:rigidity} shows the collapse of the minimum phase rigidity together with the square-root eigenvalue gap as the twist passes through the EP.

\begin{figure}[t]
\centering
\includegraphics[scale=0.33]{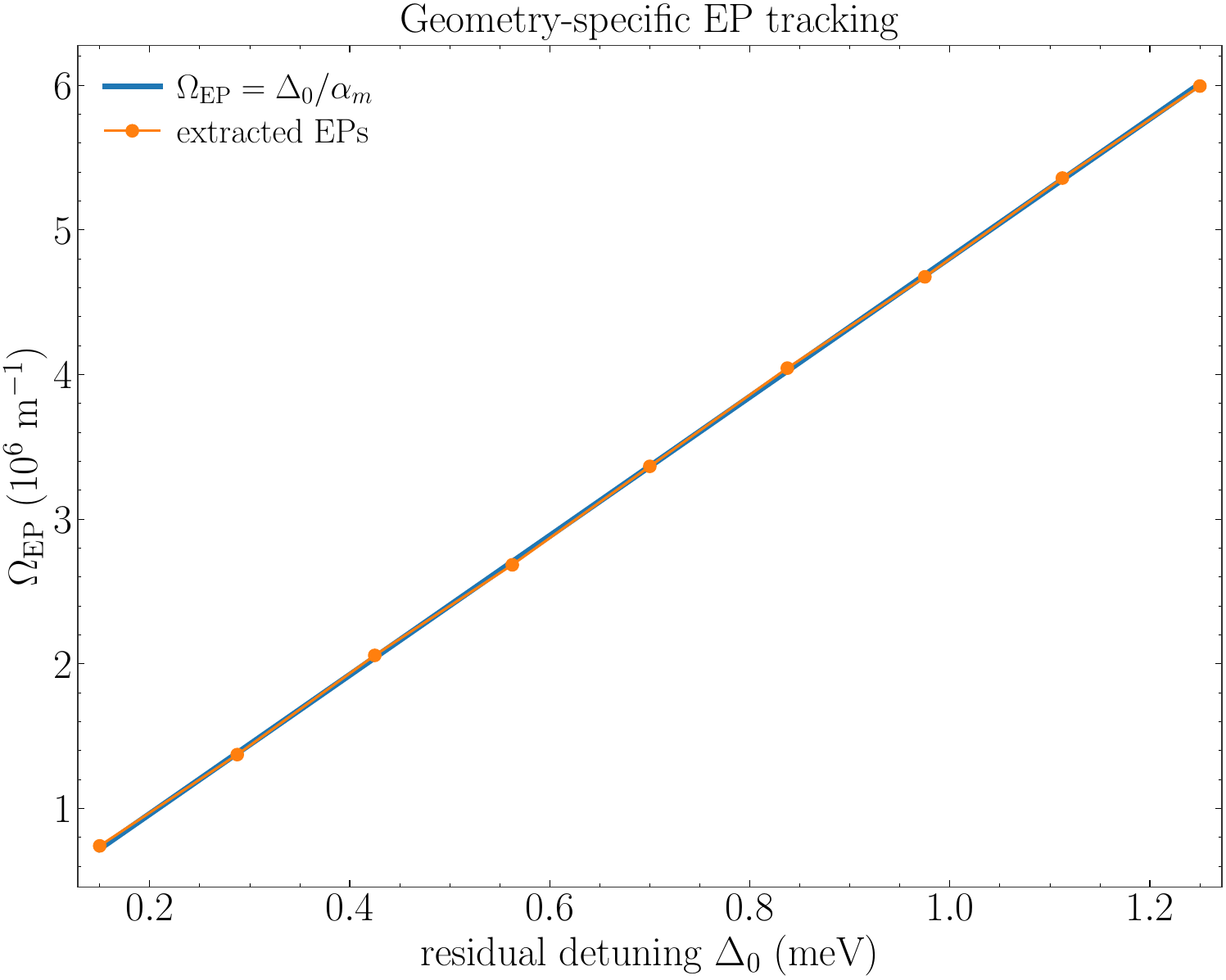}
\caption{Geometry-specific tracking of the exceptional point.
Changing the residual detuning shifts the EP linearly according to $\Omega_{\rm EP}=\Delta_0/\alpha_m$.
The points represent EPs extracted from the same two-mode condition after changing $\Delta_0$, while the solid line is the geometric prediction.}
\label{fig:tracking}
\end{figure}

The most direct geometry-specific test is not the existence of an EP, but the way its position changes as the residual detuning is varied. For fixed OAM sector, Eq.~\eqref{OmegaEPSection} predicts a straight line in the plane $(\Delta_0,\Omega_{\rm EP})$. Figure~\ref{fig:tracking} illustrates this tracking using a representative calibrated slope as a convenient conversion between detuning and twist.
In a photonic device, the same construction is obtained by replacing energy detunings with propagation-constant or cavity-frequency detunings and by using the experimentally measured OAM splitting slope. A family of EPs following this straight line would be difficult to mimic with a purely dissipative tuning parameter because the slope is fixed by the calibrated geometric OAM splitting.

\subsection{Static and dynamical geometric encircling}

The two-parameter plane in Fig.~\ref{fig:epplane} suggests a direct encircling protocol.
For the reciprocal model, define the normalized loop
\begin{equation}
\frac{\Omega(\theta)}{\Omega_{\rm EP}}
=1+\rho\cos\theta,
\qquad
\frac{\Delta\gamma(\theta)}{4g}
=1+\rho\sin\theta,
\label{loop}
\end{equation}
with $0<\rho<1$ and $0\leq\theta\leq2\pi$.
This path surrounds the EP at $(\Omega/\Omega_{\rm EP},\Delta\gamma/4g)=(1,1)$ without crossing it.
Analytically continuing the square root in Eq.~\eqref{dimensionless_lambdas} around this loop sends one eigenvalue branch into the other after a single circuit; a second circuit is required to return to the original branch. This branch exchange is a topological signature of the square-root singularity and is stronger evidence for an EP than a simple resonance enhancement.

\begin{figure*}[t]
\centering
\includegraphics[scale=0.37]{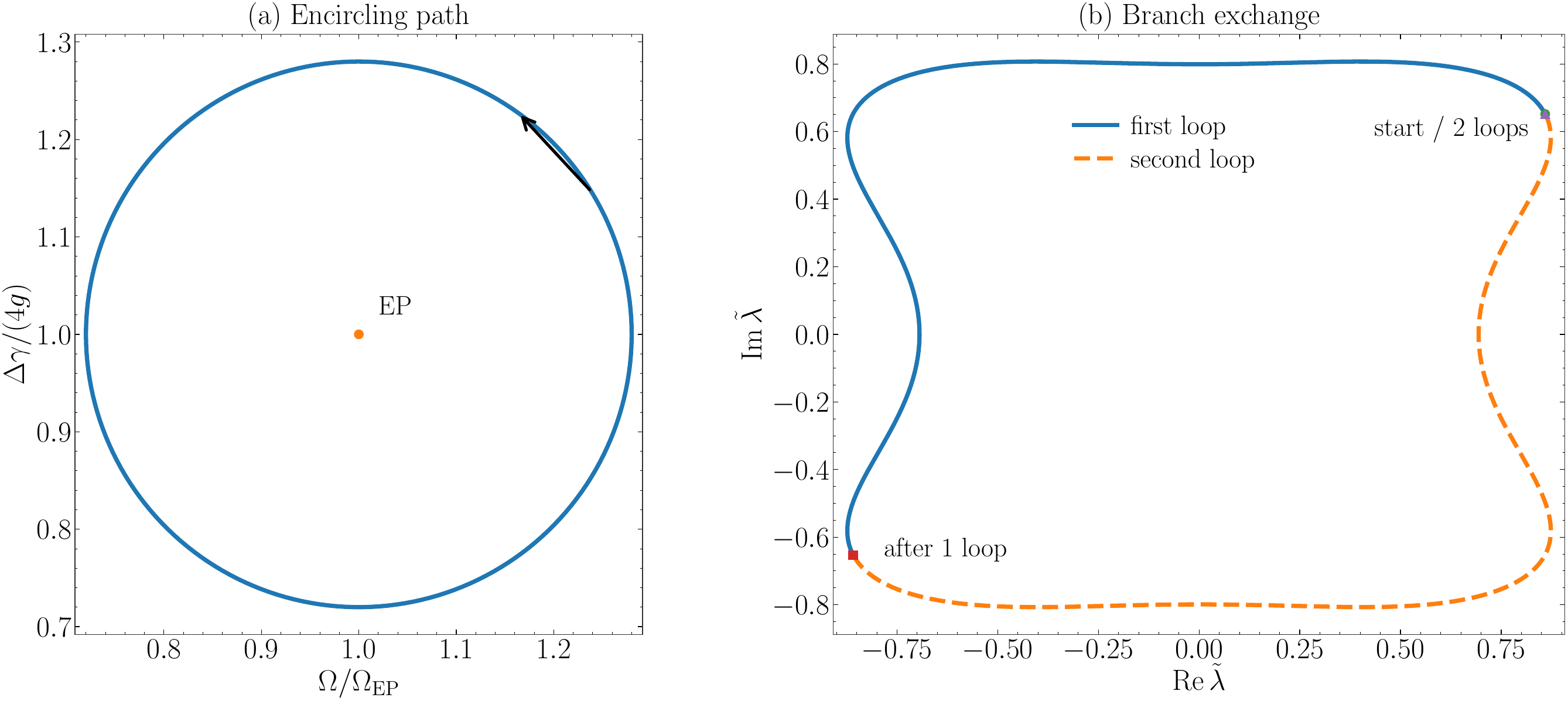}
\caption{Static geometric encircling of the exceptional point.
(a) Closed path in the parameter plane spanned by normalized twist and normalized linewidth contrast. The loop surrounds the EP while avoiding the singular point.
(b) Analytic continuation of one shifted eigenvalue branch over two circuits.
After one circuit, the branch has moved to the partner Riemann sheet; after two circuits, it returns to the starting point.}
\label{fig:encircling}
\end{figure*}

\begin{figure*}[t]
\centering
\includegraphics[scale=0.37]{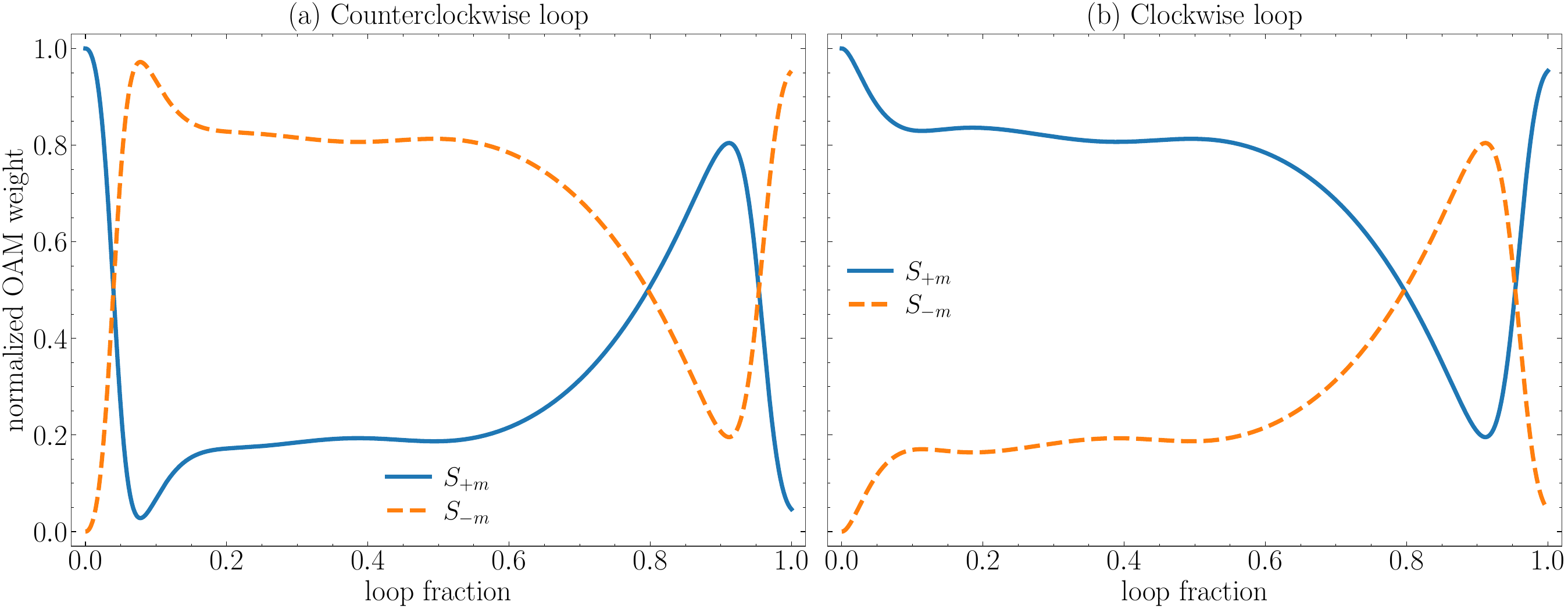}
\caption{Dynamical encircling in the two-mode OAM envelope model.
The twist and linewidth contrast follow the loop in Eq.~\eqref{loop}, but with opposite orientations in the two panels.
The normalized OAM weights show direction-dependent transfer between the two chiral modes, illustrating how a geometric loop around the EP can be converted into OAM-state dynamics.}
\label{fig:dynamic}
\end{figure*}

A propagation experiment can be extended by varying the twist and loss patterns along the propagation coordinate. In the two-mode envelope approximation, the field amplitudes obey the reduced $2\times2$ Hamiltonian $H_{\rm eff}^{(2)}$ obtained from Eq.~\eqref{HOAM} by projecting onto the target $\ell=\pm m$ doublet,
\begin{equation}
i\frac{d}{dz}
\begin{pmatrix}a_{+m}\\ a_{-m}\end{pmatrix}
=
H_{\rm eff}^{(2)}\big[\Omega(z),\Delta\gamma(z)\big]
\begin{pmatrix}a_{+m}\\ a_{-m}\end{pmatrix}.
\label{dynamic_oam}
\end{equation}
Figure~\ref{fig:dynamic} shows a representative integration for clockwise and counterclockwise loops. The plotted quantities are the normalized OAM weights $S_{+m}$ and $S_{-m}$. Because non-Hermitian adiabaticity can break down near an EP~\cite{Doppler2016,Xu2016,Qi2024}, the final OAM content depends on the loop orientation.
This orientation dependence is not used here as a precision-sensing claim; it is a dynamical signature that the geometric path has encircled a defective degeneracy.

The static and dynamical protocols test complementary aspects of the same singularity.
Static reconstruction of the complex spectrum establishes the Riemann-sheet topology, while dynamical propagation connects the geometry-controlled EP to observable OAM conversion. Together with phase-rigidity collapse and linear EP tracking, these results give the article a concrete photonic target rather than only an abstract detuning argument.

\section{Photonic implementation and falsifiable signatures}\label{secVII}

The photonic implementation is particularly natural because paraxial propagation maps the optical envelope equation onto a Schr\"odinger-like problem, with the propagation coordinate playing the role of time. Helical and femtosecond-laser-written waveguide arrays are already established platforms for engineering propagation-dependent synthetic gauge structures, topological transport, non-Hermiticity, and nonlinear dynamics~\cite{Szameit2010,Rechtsman2013,Maczewsky2020,Yan2024}. In parallel, degenerate cavities, microrings, and on-chip waveguide systems provide direct access to OAM modes and OAM-changing couplings~\cite{Hayenga2019,Yang2023,Qi2024,Forbes2024}. These ingredients make a photonic realization the most direct route to an OAM-resolved implementation of the present proposal.

In the paraxial regime, the scalar field envelope satisfies
\begin{equation}
i\frac{\partial\psi}{\partial z}
=
-
\frac{1}{2k_0n_0}\nabla_\perp^2\psi
+
V_{\rm tw}(r,\phi;z)\psi
+
V_{\rm NH}(r,\phi;z)\psi,
\label{paraxial}
\end{equation}
where $V_{\rm tw}$ encodes the helicoidal or synthetic twist and $V_{\rm NH}$ describes absorption, gain, radiative leakage, or engineered loss. Projecting Eq.~\eqref{paraxial} onto the OAM pair $|m\rangle,|-m\rangle$ gives Eq.~\eqref{Heff}, with energies replaced by propagation constants when appropriate. The geometric slope need not be postulated; it follows from a helically twisted guide. Let the transverse index perturbation rotate rigidly along the propagation axis at twist rate $\tau$ (units of rad/length), $\delta n(r,\phi,z)=\delta n_0(r,\phi-\tau z)$, so that $V_{\rm tw}=-(k_0/n_0)\,\delta n_0(r,\phi-\tau z)$. Passing to the co-rotating azimuth $\tilde\phi=\phi-\tau z$, in which the structure is stationary, the envelope $\chi(r,\tilde\phi,z)\equiv\psi(r,\phi,z)$ obeys
\begin{equation}
i\frac{\partial\chi}{\partial z}
=\left[-\frac{1}{2k_0n_0}\nabla_\perp^2+V_{\rm tw}(r,\tilde\phi)-\tau\,\hat{\ell}+V_{\rm NH}\right]\chi,
\label{rotating_frame}
\end{equation}
where $\hat\ell=-i\partial_{\tilde\phi}$ is the OAM operator. The rotation generates exactly an OAM-linear term $-\tau\hat\ell$, the optical analog of the chiral Zeeman shift $-\hbar\omega_s m$ of Sec.~\ref{secII}. An OAM eigenmode $\ell$ therefore acquires a propagation-constant shift $-\tau\ell$, and the opposite-OAM doublet acquires the geometric detuning
\begin{equation}
\Delta\beta_{\rm geom}(\tau)=\beta_{+m}-\beta_{-m}=-2m\tau,
\qquad
\frac{\partial\,\Delta\beta_{\rm geom}}{\partial\tau}=-2m .
\label{slope_photonic}
\end{equation}
This reproduces Eq.~\eqref{DeltaGeom} under the identification $\hbar\omega_s\leftrightarrow\tau$: the combination $\Omega k_z$ that sets the chiral scale in the massive problem is replaced by the directly measurable geometric twist rate of the guide. The higher-order terms generated by the rotating frame ($\propto\tau^2 r^2$ and spin-orbit corrections) are diagonal in $|\ell|$ and do not lift the $\pm m$ degeneracy at leading order; they renormalize the radial confinement and $\Delta_0$ rather than the slope. The geometric slope is thus fixed by the twist rate and is independent of the non-Hermitian parameters, a property exploited throughout this work. In the optical platform the abstract twist $\Omega$ of Secs.~\ref{secII}--\ref{secVI} is therefore realized as the physical guide twist rate $\tau$ at a fixed axial sector, so that the swept coordinate $\Omega/\Omega_{\rm EP}$ and the compensation point $\Omega_{\rm EP}$ correspond directly to $\tau/\tau_{\rm EP}$ and $\tau_{\rm EP}=\Delta\beta_0/2m$.
The OAM coupling can be created by an angular grating proportional to $\cos(2m\phi)$, by an elliptic boundary deformation, or by an index modulation carrying angular momentum $2m$.
The linewidth contrast can be introduced by an azimuthally structured absorber, an OAM-selective outcoupler, or a gain profile whose overlap differs for the two chiral modes.

\subsection{Experimental workflow}

A concrete experimental workflow is as follows.

\begin{enumerate}
\item \emph{Mode preparation.}
Launch or excite the two OAM modes $|m\rangle$ and $|-m\rangle$ and verify that other OAM components are weak over the propagation or cavity-lifetime scale.
The OAM weights are obtained from
\begin{align}
F_\ell(r,z)&=\frac{1}{2\pi}\int_0^{2\pi}d\phi\,
\psi(r,\phi,z)e^{-i\ell\phi},
\nonumber\\
S_\ell(z)&=\int_0^{R_{\rm out}}dr\,r\,|F_\ell(r,z)|^2 .
\label{OAMweights}
\end{align}

\item \emph{Hermitian calibration.}
With gain and loss minimized, measure the two real propagation constants or resonance frequencies as functions of the effective helicoidal parameter.
A linear fit gives $\Delta_0$ and the geometric slope $\alpha_m$.
This step is essential because it separates the geometric control from non-Hermitian broadening.

\item \emph{Coupling calibration.}
Turn on the angular perturbation that couples $|m\rangle$ and $|-m\rangle$.
In the nearly-Hermitian limit, the avoided-crossing gap at zero detuning equals $2|g|$.
This provides the target linewidth contrast $|\Delta\gamma|=4|g|$ for the reciprocal real-coupling EP.

\item \emph{Loss engineering.}
Introduce OAM-selective loss or gain until the measured linewidth difference approaches the target value. This can be done by moving an absorber, changing the pump overlap, or opening a radiative channel with different overlap for the two OAM modes.

\item \emph{EP localization.}
Sweep $\Omega$ through the compensation point while monitoring the complex propagation constants. The EP is identified by the simultaneous coalescence of the real splitting, the linewidth splitting, and the eigenmode profile.

\item \emph{Eigenvector-sensitive verification.}
Measure the overlap or phase rigidity of the two modes.
For right and left eigenvectors, a useful diagnostic is
\begin{equation}
r_j=
\frac{|\langle\psi_j^L|\psi_j^R\rangle|}
{\sqrt{\langle\psi_j^L|\psi_j^L\rangle\langle\psi_j^R|\psi_j^R\rangle}},
\label{phase_rigidity}
\end{equation}
which collapses at an ideal second-order EP.
Mode-profile coalescence or branch exchange under encircling provides an equivalent test when left eigenvectors are difficult to reconstruct.

\item \emph{Geometric encircling.}
Implement the loop in Eq.~\eqref{loop} either by static scanning or by adiabatically varying the twist and loss along the propagation direction.
The expected outcome is sheet exchange for the static spectrum and, in dynamical propagation, direction-dependent OAM conversion analogous to known EP-encircling experiments~\cite{Doppler2016,Xu2016,Qi2024}.
\end{enumerate}

This workflow makes the proposal operational.
The claim is not that any resonance enhancement near a twisted guide proves an EP.
Rather, the evidence consists of a calibrated real geometric slope, a non-Hermitian linewidth balance, eigenvalue and eigenvector coalescence, and branch exchange under a loop whose horizontal coordinate is the helicoidal twist.

\subsection{Calibration and falsifiable signatures}

The proposal is experimentally meaningful only if the exceptional point can be identified as geometry-controlled rather than merely as an ordinary non-Hermitian crossing produced by losses.
A practical calibration protocol is therefore to first measure the real two-mode detuning in a weakly open or nearly Hermitian configuration,
\begin{equation}
\Delta(\Omega)=\Delta_0-\alpha_m\Omega,
\qquad
\alpha_m=\frac{2m\hbar^2\pi n_z}{m^\ast L},
\label{calibration_slope}
\end{equation}
and then tune the linewidth contrast and coupling until the pole coalescence occurs at the predicted compensation point. In photonic language, the same equation applies to propagation-constant detunings, with energies replaced by the corresponding optical propagation constants.

This calibration leads to three falsifiable predictions. First, if the residual detuning $\Delta_0$ is shifted by an external gate, strain field, refractive-index offset, or boundary deformation, the exceptional-point twist should move linearly according to
\begin{equation}
\frac{d\Omega_{\rm EP}}{d\Delta_0}=\frac{1}{\alpha_m}.
\end{equation}
Second, reversing the helicoidal handedness reverses the sign of the chiral splitting and interchanges the ordering of the $m$ and $-m$ modes, while leaving the radial confinement scale unchanged. Third, comparing different OAM or axial sectors should reveal the scaling $\alpha_m\propto m n_z/L$. These are geometry-specific signatures and cannot be reproduced by a purely dissipative tuning parameter alone.

It is worth stating precisely what makes the control geometric rather than a generic real-detuning parameter. Any mechanism producing an OAM-linear detuning (mechanical rotation, a chirped azimuthal grating, or a synthetic magnetic field) could in principle move an EP; what singles out the helicoidal mechanism is that (i) the slope equals an independently measured geometric quantity, the twist rate $\tau$ of a helically twisted guide (derived in the paraxial mapping above), or equivalently the number of helical turns per unit length, and (ii) it inherits the discrete scaling $\alpha_m\propto m\,n_z/L$ and the sign reversal under handedness inversion, neither of which a purely dissipative parameter exhibits. The accessible subset of these tests is platform-dependent: twisted-fiber and femtosecond-written helical-waveguide platforms give direct access to $\tau$ and to handedness reversal, whereas microring and degenerate-cavity platforms most naturally expose the linear $\Omega_{\rm EP}(\Delta_0)$ tracking and the $m$-scaling at fixed geometry. Reporting at least two independent signatures from this list distinguishes geometric control from incidental loss tuning.

The same analysis also clarifies what would not constitute evidence for the proposed mechanism. Varying only $\Omega$ at a linewidth contrast different from $|\Delta\gamma|=4|g|$ generally gives an avoided crossing or a linewidth crossing, not a second-order exceptional point. Likewise, observing an enhanced response near a resonance is insufficient unless accompanied by square-root branch behavior, eigenvector coalescence, or the collapse of phase rigidity. Perturbations that preserve the two-mode structure mainly renormalize $\Delta_0$, $\alpha_m$, $g$, and $\Delta\gamma$. Perturbations that couple the selected pair strongly to additional nearby modes require a multimode extension, in which the same geometric slope can still move the system through higher-order or hybrid exceptional degeneracies.

\section{Device-level paraxial validation}
\label{secVIII}

The coupled-mode arguments above isolate the geometry-controlled exceptional point in an effective two-mode language. To show that the mechanism is not an artifact of that reduction, we now extract a geometry-controlled exceptional point directly from a paraxial device model whose parameters are not imposed but computed from a concrete refractive-index structure. We deliberately treat two physically distinct realizations, one reciprocal and one asymmetric-coupling Wiersig-type platform, and show that the helical twist drives both to a square-root singularity of the same character.

\subsection{Paraxial coupled-mode model}

We consider an annular OAM waveguide with an index profile
$n(r)=n_0+\Delta n\,e^{-(r-R)^2/2w^2}$, taken here as a silicon-nitride-like ring
($n_0=2.0$, $R=5~\mu$m, $w=0.5~\mu$m, $\Delta n=0.05$) at $\lambda=1.55~\mu$m.
For each azimuthal order $\ell$, the radial paraxial eigenproblem
\begin{equation}
\Big[-\tfrac{1}{2k_0n_0}\big(\partial_r^2+\tfrac1r\partial_r-\tfrac{\ell^2}{r^2}\big)
-\tfrac{k_0}{n_0}\Delta n(r)\Big]\mathcal R_\ell=\beta^{0}_\ell\,\mathcal R_\ell
\end{equation}
is solved on a radial grid, giving the unperturbed propagation constants
$\beta^{0}_\ell$ (degenerate in $\pm\ell$) and radial modes $\mathcal R_\ell(r)$.
Projecting the full paraxial operator onto the angular harmonics $e^{i\ell\phi}$
(with $|\ell|\le L$) and passing to the co-rotating frame of Eq.~\eqref{rotating_frame}, the device Hamiltonian becomes the multimode matrix
\begin{equation}
H_{\ell\ell'}=\big(\beta^{0}_\ell-\tau\ell\big)\delta_{\ell\ell'}
+\widehat V_{\ell-\ell'},
\label{Hdevice}
\end{equation}
where the twist enters exactly as the OAM-linear term $-\tau\ell$ derived in
Sec.~\ref{secVII} and $\widehat V_{k}$ are the angular Fourier components of the index and loss perturbations, weighted by the computed radial overlaps. The target doublet is $\ell=\pm m$ with $m=1$; the nearest spectator OAM mode sits $2.97~\mathrm{rad/mm}\approx 6|g|$ away, so the doublet is spectrally isolated.

\subsection{Two physical realizations}

\emph{Reciprocal platform.} A reciprocal index grating $\propto\cos(2m\phi)$
couples $\ell=\pm m$ with a real, symmetric amplitude $g_{12}=g_{21}=g$, while an
OAM-selective (helicity-selective) radiative outcoupler~\cite{Forbes2024,Yang2023}
imposes a tunable, diagonal loss contrast $\gamma_{+m}-\gamma_{-m}=\Delta\gamma$.
This realizes the reciprocal exceptional point of Sec.~\ref{secIV}, expected at
$\Delta=0$, $|\Delta\gamma|=4|g|$.

\emph{Asymmetric-coupling (Wiersig-type) platform.} Two finite-size scatterers (one
absorptive) at different azimuths produce asymmetric backscattering between the
co- and counter-rotating modes, $g_{12}\neq g_{21}$ (complex)~\cite{Wiersig2014,Wiersig2020}, realizing the general complex-coupling
case of Sec.~\ref{secIV}. A weakly OAM-selective loss serves as the non-Hermitian parameter. Here, the exceptional point is expected at a \emph{nonzero} real detuning.

In both platforms, the exceptional point is located by tuning the twist $\tau$
(real parameter) together with the openness parameter, and we verify coalescence on the
\emph{full} multimode matrix~\eqref{Hdevice}, not on a reduced $2\times2$ block.

\subsection{Results}

\begin{figure*}[t]
\centering
\includegraphics[scale=0.42]{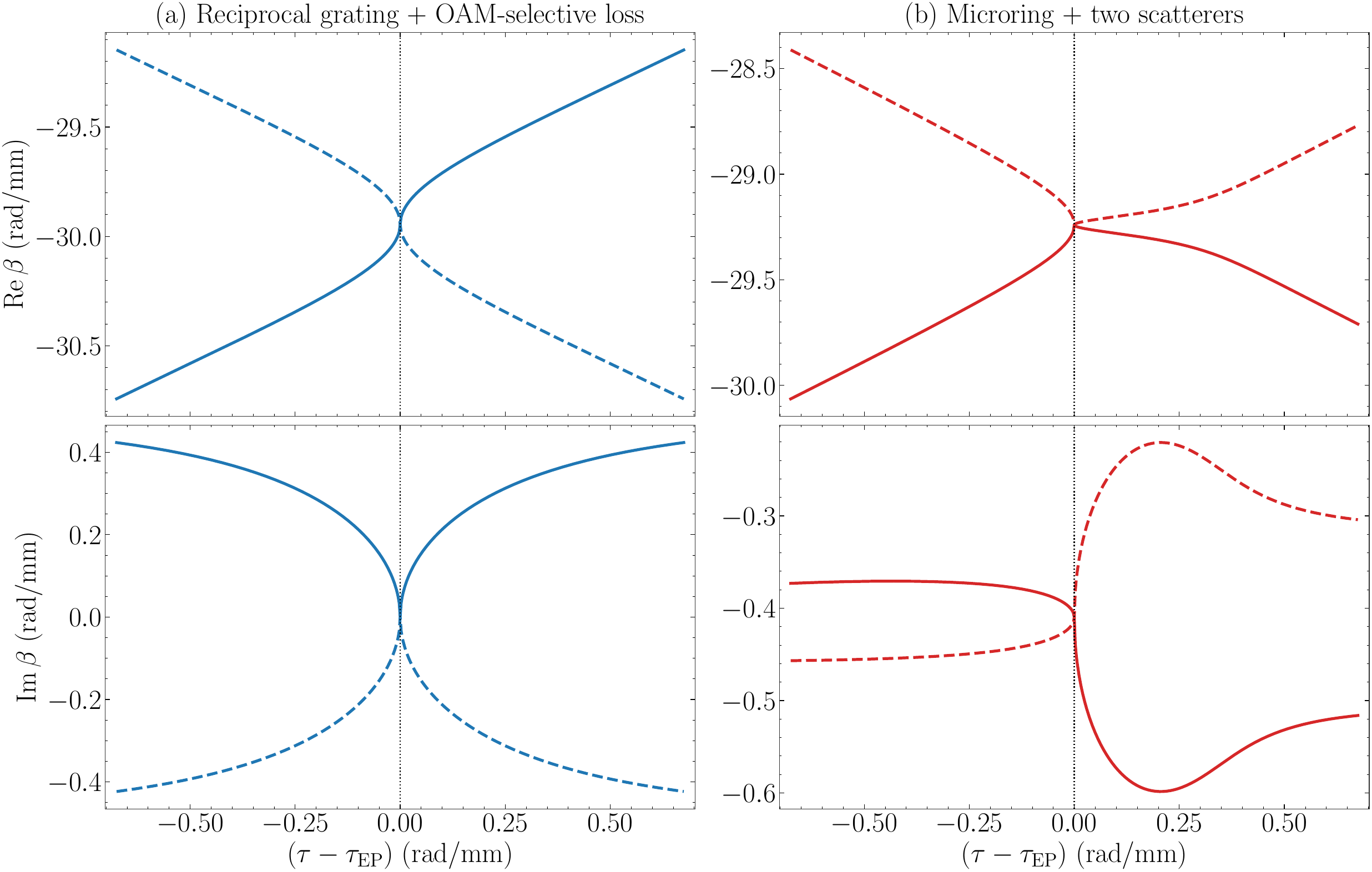}
\caption{Device-level eigenvalue coalescence. Real (top) and imaginary (bottom)
parts of the two target propagation constants versus normalized twist for the
reciprocal grating platform (left) and the asymmetric-scatterer platform
(right), computed from the full multimode model~\eqref{Hdevice}. Both branches
coalesce at $\tau=\tau_{\rm EP}$.}
\label{fig:sim_eigtrack}
\end{figure*}

\begin{figure*}[t]
\centering
\includegraphics[width=\textwidth]{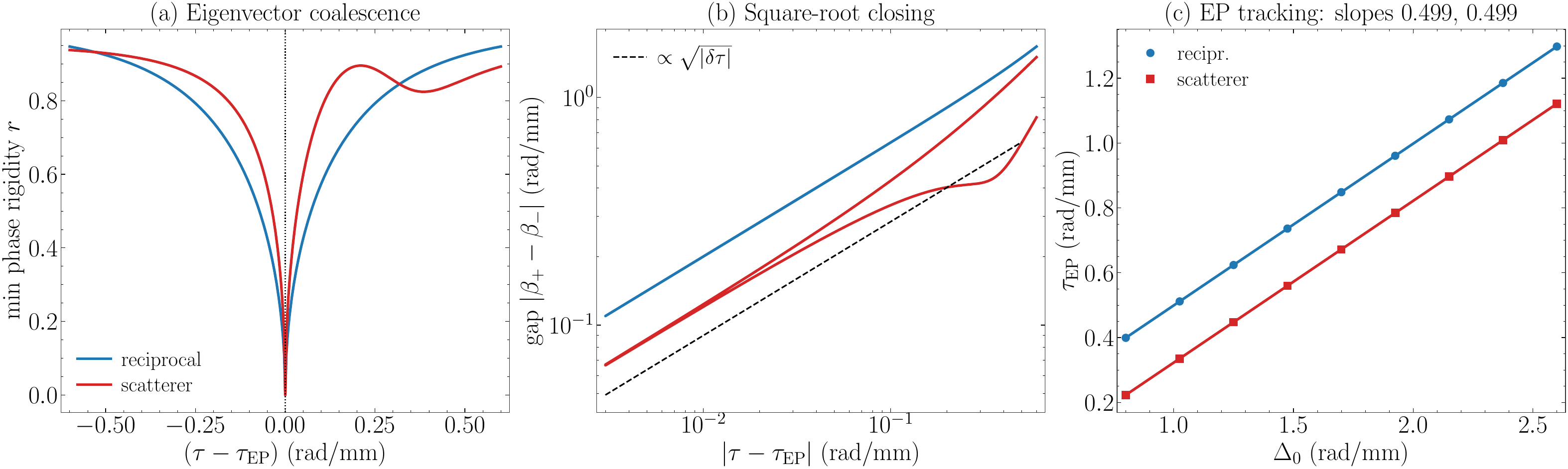}
\caption{Falsifiable signatures from the device model.
(a) Minimum phase rigidity collapses at the exceptional point for both platforms
(eigenvector coalescence).
(b) Eigenvalue gap versus detuning from the EP on a log-log scale, following the
$\sqrt{|\tau-\tau_{\rm EP}|}$ law (dashed).
(c) Linear tracking of the EP twist with residual detuning $\Delta_0$; the fitted
slopes ($0.499$ for both platforms) match the geometric prediction $1/(2m)$.}
\label{fig:sim_signatures}
\end{figure*}

\begin{figure*}[t]
\centering
\includegraphics[scale=0.42]{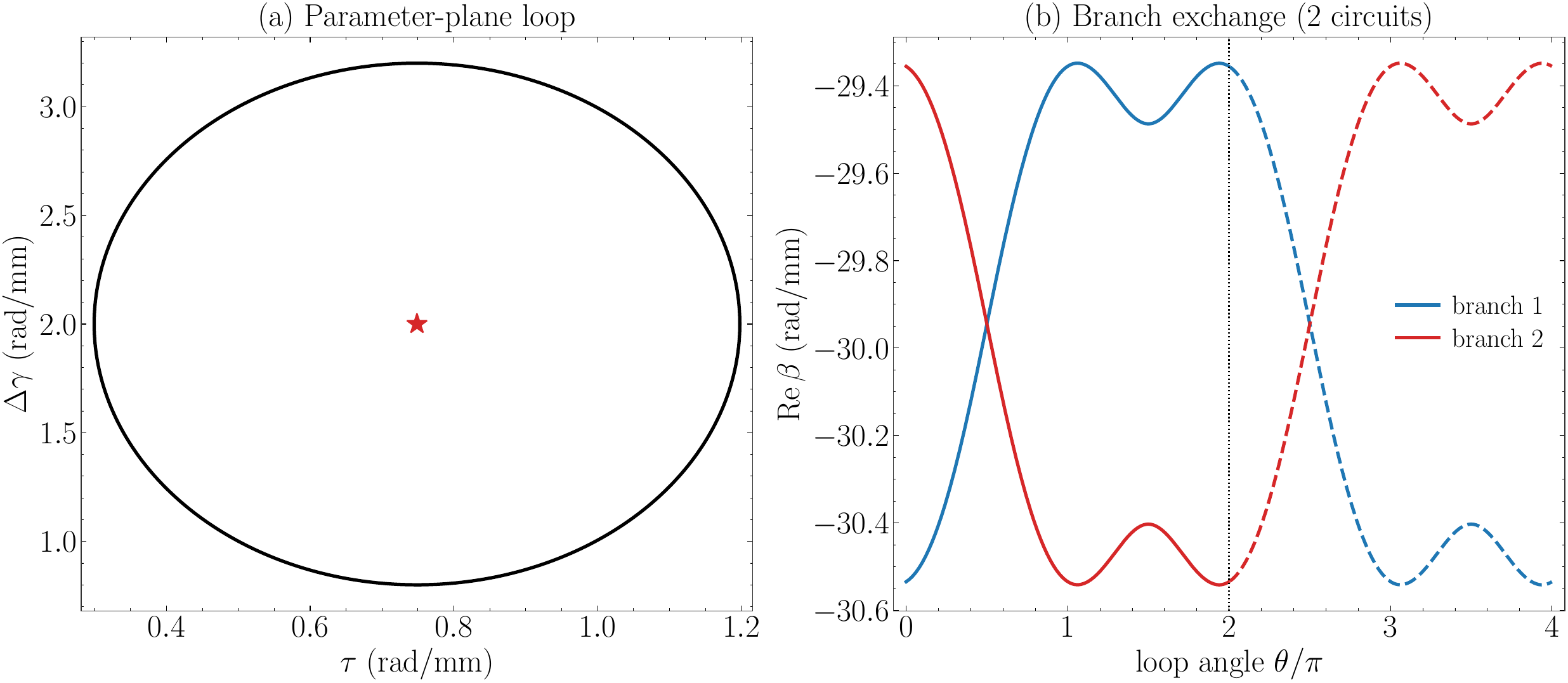}
\caption{Geometric encircling in the device model (reciprocal platform).
(a) Closed loop in the $(\tau,\Delta\gamma)$ plane surrounding the exceptional
point (star). (b) Analytic continuation of the two device eigenvalues over two
circuits: the branches exchange after one circuit and return after two, the
Riemann-sheet signature of the square-root singularity.}
\label{fig:sim_encircling}
\end{figure*}

Figure~\ref{fig:sim_eigtrack} shows the two target propagation constants as the
twist is swept through the compensation point: in both platforms, the real and
imaginary parts coalesce at $\tau=\tau_{\rm EP}$, the hallmark of a second-order
exceptional point. Figure~\ref{fig:sim_signatures}(a) shows that the minimum
phase rigidity collapses to $\sim10^{-7}$ there, confirming eigenvector (not
merely eigenvalue) coalescence, while panel~(b) confirms the non-analytic
$|\beta_+-\beta_-|\propto\sqrt{|\tau-\tau_{\rm EP}|}$ closing. Panel~(c) is the
central geometry-specific test: re-locating the exceptional point for a family of
residual detunings $\Delta_0$ gives a straight line $\tau_{\rm EP}(\Delta_0)$ with
fitted slope $0.499$ for both platforms, in agreement with the predicted
$1/(2m)=0.5$. Finally, Fig.~\ref{fig:sim_encircling} implements the
parameter-plane loop of Eq.~\eqref{loop}: analytic continuation of the device
eigenvalues exchange the two branches after one circuit and restore them after two, the Riemann-sheet signature of the square-root branch point.

\begin{table*}[t]
\caption{Exceptional-point parameters \emph{extracted} from the paraxial device
model of Eq.~\eqref{Hdevice} (not imposed), in propagation-constant units (rad/mm). Residual detuning $\Delta_0=1.5~\mathrm{rad/mm}$, target coupling scale $|g|\!\approx\!0.5~\mathrm{rad/mm}$. Both platforms reach a genuine second-order EP (phase rigidity $r_{\min}\!\to\!0$); the idealized two-mode condition is essentially exact for the reciprocal platform and acquires a multimode correction for the asymmetric-coupling one.}
\label{tab:simparams}
\begin{ruledtabular}
\begin{tabular}{lcc}
quantity & Reciprocal & Asymmetric-coupling (Wiersig type)\\
\colrule
$\tau_{\rm EP}$ (rad/mm)            & $0.749$ & $0.572$\\
$\Delta\gamma_{\rm EP}$ (rad/mm)    & $2.00$  & $0.297$\\
$\Delta_{\rm EP}$ (rad/mm)          & $\simeq 0$ & $0.353$\\
$|g_{12}|,\,|g_{21}|$ (rad/mm)      & $0.50,\,0.50$ & $0.125,\,0.325$\\
coupling asymmetry $|g_{12}/g_{21}|$& $1.00$  & $0.38$\\
$|\Delta\gamma|/4|g|$               & $1.000$ & --- \\
EP tracking slope $d\tau_{\rm EP}/d\Delta_0$ & $0.499$ & $0.499$\\
min.\ phase rigidity $r_{\min}$     & $9\times10^{-8}$ & $3\times10^{-7}$\\
\end{tabular}
\end{ruledtabular}
\end{table*}

Table~\ref{tab:simparams} collects the extracted parameters. For the reciprocal platform, the device reproduces the analytic conditions to numerical precision:
$\Delta_{\rm EP}\simeq0$ and $|\Delta\gamma|/4|g|=1.000$. For the asymmetric-coupling
platform the exceptional point occurs at a finite real detuning $\Delta_{\rm EP}=0.353~\mathrm{rad/mm}$ with strongly asymmetric coupling $|g_{12}/g_{21}|=0.38$, exactly the regime in which Sec.~\ref{secIV} predicts an EP without requiring $\Delta=0$. We verified that $\tau_{\rm EP}$ is converged in the angular window for $|\ell|\le4$ (changes below $0.1\%$ on extending to $|\ell|\le7$). Evaluated on the reduced doublet, the idealized two-mode degeneracy condition, Eq.~\eqref{EP_general}, is met with a relative residual $\sim10^{-5}$ for the reciprocal platform but only $\sim10^{-1}$ for the asymmetric-coupling one: the broadband asymmetric coupling induces a genuine multimode correction, yet the full device still exhibits a clean second-order EP. This is precisely the content of the multimode-robustness analysis of Sec.~\ref{secVI}: the geometric slope continues to steer the open system to a defective degeneracy, while the precise location, for asymmetric coupling, must be read from the full mode network rather than from
the bare $2\times2$ block.

The parameters of Table~\ref{tab:simparams} sit in the operating window quoted in
Sec.~\ref{secIX} ($|g|$ of order $0.5~\mathrm{rad/mm}$, $\tau_{\rm EP}$ of order
$1~\mathrm{rad/mm}$, well within twisted-fiber and helical-waveguide
capabilities~\cite{Szameit2010,Rechtsman2013}), so the geometry-controlled exceptional point is not only analytically consistent but also numerically reachable in a realistic OAM photonic device.

\section{OAM-resolved spectral response and 3D EP surfaces}\label{secIX}

The step-by-step protocol of Sec.~\ref{secVII} can be translated directly into observables of a cavity or ring-resonator experiment.
A minimal measurement-level description is obtained by coupling the effective OAM doublet to external excitation and collection channels.
Let $\mathbf{s}_{\rm in}$ and $\mathbf{s}_{\rm out}$ denote the input and output channel amplitudes, and let $W$ be the matrix that couples those channels to the two OAM modes.
The standard input-output response is then
\begin{equation}
\mathbf{s}_{\rm out}(\omega)
=
\mathbf{s}_{\rm in}
-
iW^\dagger
\left[\omega I_2-H_{\rm eff}(\Omega)\right]^{-1}
W\mathbf{s}_{\rm in},
\label{inputoutput}
\end{equation}
where $I_2$ is the identity in the two-dimensional OAM doublet subspace. For consistency, the channel coupling $W$ and the radiative linewidth matrix are not independent: energy conservation in the background fixes $\Gamma^{\rm(rad)}=WW^\dagger$, so that $H_{\rm eff}$ contains $-i\Gamma^{\rm(rad)}/2$ in addition to any internal nonradiative loss. The diagonal linewidths $\gamma_{1,2}$ entering Eq.~\eqref{Heff} should therefore be understood as the sum of radiative and internal contributions.
The intracavity OAM amplitudes are
\begin{equation}
\mathbf{a}(\omega,\Omega)
=
\left[\omega I_2-H_{\rm eff}(\Omega)\right]^{-1}
W\mathbf{s}_{\rm in}.
\label{intracavity}
\end{equation}
From these quantities one extracts the total spectral response $I(\omega,\Omega)=\mathbf{a}^\dagger\mathbf{a}$ and the OAM contrast
\begin{equation}
C_{\rm OAM}(\omega,\Omega)
=
\frac{|a_{+m}|^2-|a_{-m}|^2}{|a_{+m}|^2+|a_{-m}|^2}.
\label{OAMcontrast}
\end{equation}
These observables are directly connected to transmission spectroscopy, resonant outcoupling, or OAM-resolved camera images.

\subsection{Spectral response and Riemann-surface visualization}

\begin{figure*}[t]
\centering
\includegraphics[scale=0.45]{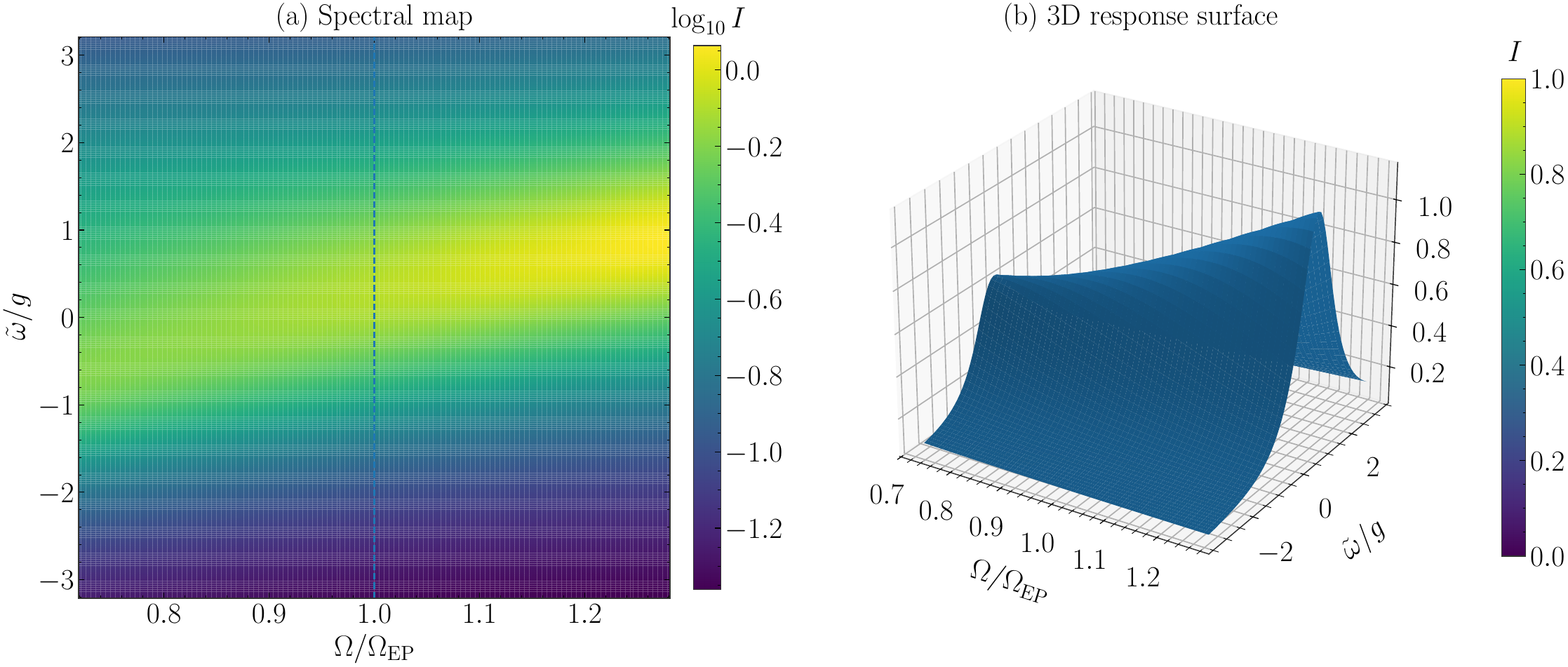}
\caption{OAM-resolved spectral response of a cavity implementation.
(a) Density plot of the normalized intracavity response $I(\tilde\omega,\Omega)$ as a function of normalized frequency detuning $\tilde\omega$ and normalized twist.
Two resonant features approach one another and merge at $\Omega=\Omega_{\rm EP}$.
(b) Three-dimensional rendering of the same response surface. This figure translates the geometry-controlled EP from the language of eigenvalues into a spectroscopy-level observable.}
\label{fig:response3d}
\end{figure*}

Figure~\ref{fig:response3d} shows the response predicted by Eqs.~\eqref{inputoutput} and \eqref{intracavity} for a reciprocal EP with equal excitation of the two OAM channels.
The left panel shows a conventional density plot of the intracavity response versus the normalized frequency and the normalized twist. The right panel is a three-dimensional rendering of the same quantity. The two resonant ridges move together and merge at the geometry-controlled EP. This is the response-level version of the eigenvalue coalescence discussed in Secs.~\ref{secIV} and \ref{secVI}: rather than looking directly at the poles, one sees the same singularity in an experimentally accessible spectral map.

\begin{figure*}[t]
\centering
\includegraphics[scale=0.45]{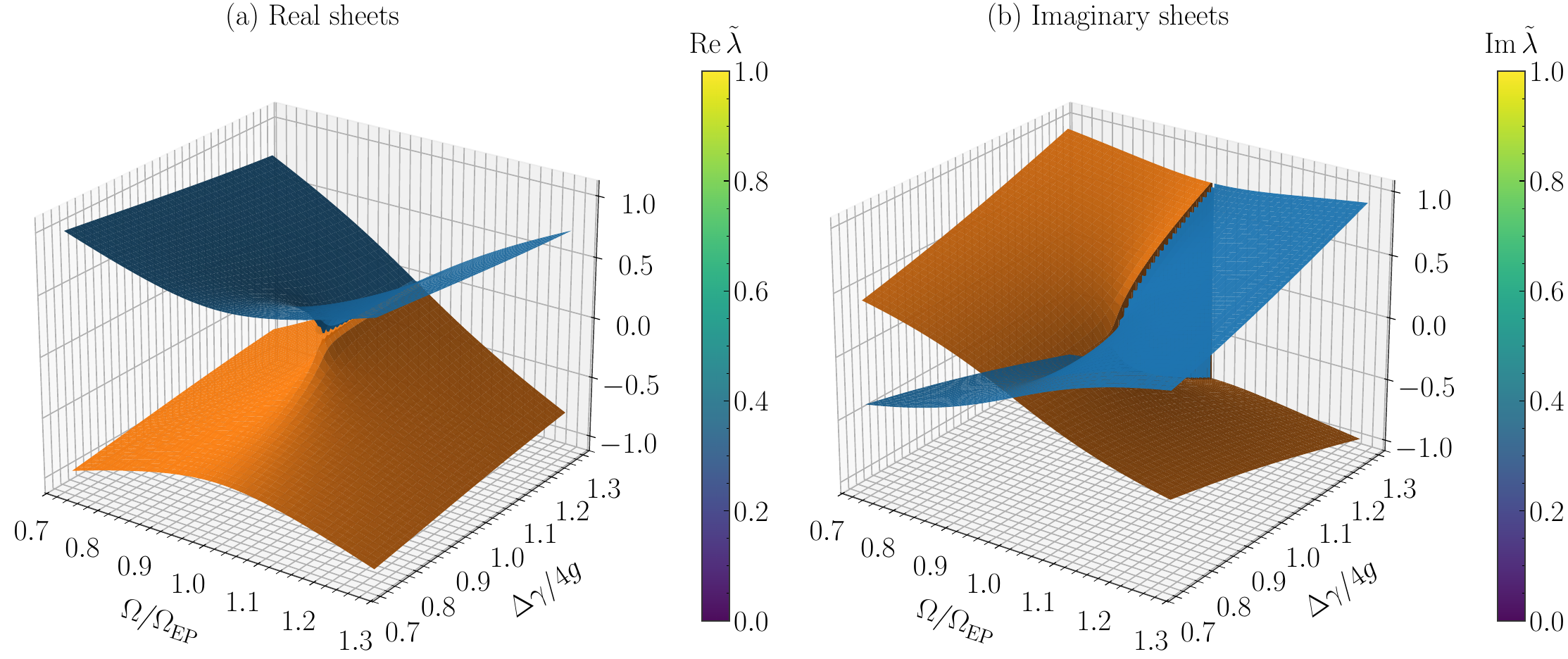}
\caption{Three-dimensional visualization of the EP topology.
(a) Real parts of the two eigenvalue sheets and (b) imaginary parts of the same sheets over the plane $(\Omega/\Omega_{\rm EP},\Delta\gamma/4g)$.
The two sheets meet at the geometry-controlled EP, making the square-root branch structure explicit.
These surfaces are the three-dimensional counterpart of the two-dimensional EP-condition plot in Fig.~\ref{fig:epplane}.}
\label{fig:riemann3d}
\end{figure*}

A second set of plots emphasizes the branch-point topology itself.
Figure~\ref{fig:riemann3d} renders the two eigenvalue sheets in three dimensions over the parameter plane spanned by normalized twist and normalized linewidth contrast.
The surfaces meet at the same point highlighted in Fig.~\ref{fig:epplane}, but the three-dimensional view makes the square-root sheet structure more transparent.
Such a plot is not itself an experimental observable; its role is explanatory.
It shows why the static loop in Fig.~\ref{fig:encircling} exchanges sheets after one turn and provides a visually direct representation of the singularity around which the photonic protocol is designed.

The main point of this section is methodological.
The theory is no longer expressed only in terms of an effective pole condition.
It is connected to quantities routinely measured in photonic experiments: transmission or reflection maps, OAM-resolved intensity distributions, extracted linewidths, and changes in spectral topology under static or dynamical parameter loops. At the same time, the model remains intentionally lightweight. It is a coupled-mode and input-output description parameterized by the measured OAM splitting slope, not a substitute for a full-wave simulation of a specific fabricated device.

\subsection{Photonic observables and experimentally accessible scales}

For a photonic implementation, the most important experimentally accessible quantities are the real mode splitting, the linewidth difference, the avoided-crossing gap, the OAM-resolved weights, the spectral map of Fig.~\ref{fig:response3d}, and the final-state conversion probability after dynamical encircling. The geometry-controlled claim is strongest when these quantities are measured in a fixed OAM sector while $\Delta_0$ is deliberately varied. The EP should then move according to the linear law of Fig.~\ref{fig:tracking}, with the slope independently calibrated from the nearly Hermitian OAM splitting.

The present theory does not require a unique microscopic platform. What it requires is an OAM doublet whose real splitting can be tuned through a calibrated helicoidal or synthetic-helicoidal control parameter and whose linewidths can be made unequal in a controllable way. Degenerate cavities, microring or microdisk resonators with OAM-selective scattering, synthetic OAM-loop architectures, and femtosecond-written waveguide systems all satisfy these requirements in different ways~\cite{Hayenga2019,Yang2023,Qi2024,Forbes2024,Szameit2010,Rechtsman2013,Maczewsky2020,Yan2024}.
In all such cases, the coupled-mode parameters are extracted from the device rather than imposed a priori. This is precisely why the measurement-level response of Sec.~\ref{secIX} is useful: it gives a device-agnostic way to compare the experiment with the theory.

A realistic target regime is therefore not specified by absolute universal numbers but by inequalities.
The geometry-induced detuning sweep across the accessible range of $\Omega$ must be comparable to or larger than the coupling scale $g$.
The chosen OAM pair must remain spectrally isolated from spectator modes, as in Fig.~\ref{fig:multimode}.
The linewidth contrast must be tunable through the window $|\Delta\gamma|\sim 4|g|$ for the reciprocal EP.
Finally, the spectrometer or imaging setup must resolve either the pole coalescence itself or one of the stronger diagnostics already identified above: phase-rigidity collapse, OAM-profile coalescence, or loop-direction-dependent mode conversion.
When these conditions are met, the response surfaces of Figs.~\ref{fig:response3d} and \ref{fig:riemann3d} become realistic experimental targets rather than abstract illustrations.

As a concrete operating point, consider an $m=1$ OAM doublet in a femtosecond-laser-written helical waveguide array, coupled by a $q=2$ angular index modulation. A coupling with beat length $L_c=\pi/(2|g|)\approx3$~mm corresponds to $|g|\approx0.5$~rad/mm; the reciprocal EP then requires a differential loss $|\Delta\gamma|=4|g|\approx2$~rad/mm between the two chiral modes, attainable with an OAM-selective absorber or asymmetric outcoupling. With a residual detuning $\Delta\beta_0\approx4$~rad/mm set by boundary ellipticity, the compensation twist is $\tau_{\rm EP}=\Delta\beta_0/2m\approx2$~rad/mm, well within the range of twisted photonic-crystal fibers and helical waveguide arrays (typically up to $\sim10$~rad/mm). A device length of $10$--$20$~mm spans several coupling lengths and resolves the square-root coalescence. In cavity-frequency language, the same ratios read $g/2\pi$ of a few GHz, $\Delta\gamma/2\pi=4g/2\pi$, and linewidths set by a loaded quality factor $Q\sim10^4$--$10^5$ at optical frequencies, all standard in microring and degenerate-cavity OAM platforms. These numbers turn the inequalities above into a falsifiable target.

\section{Scope, limitations, and outlook}\label{secX}

The central claim of this work remains deliberately limited.
The helicoidal metric controls the real part of the detuning between chiral modes; it does not by itself generate the non-Hermitian terms required for an exceptional point.
Those terms must come from openness, gain, absorption, radiative leakage, dissipative coupling, or postselected no-jump dynamics.
Accordingly, the proposed mechanism should be tested by calibrating four effective quantities: the geometric slope $\alpha_m$, the residual detuning $\Delta_0$, the linewidth contrast $\Delta\gamma$, and the intermodal coupling $g$.

A first limitation is that the present photonic treatment, although supplemented by device-level paraxial validation, is neither a full-vector Maxwell simulation nor a finite-element optimization of a specific fabricated structure. This is intentional. The goal of the article is to isolate the geometry-controlled mechanism, identify its falsifiable signatures, and provide experimentally readable observables and realistic parameter regimes. A next step for a fully device-specific study would be to extract the same effective quantities from a full-vector electromagnetic or finite-element model of a concrete cavity or waveguide geometry and then compare them with the compact mode-space and input-output description used here.

A second limitation is the isolated-doublet assumption. If additional OAM modes lie within the scale set by $g$, $\gamma_i$, or the detuning sweep, the same helicoidal slope can still move the system through non-Hermitian singularities, but the resulting structure may be a multimode or higher-order EP rather than the square-root EP described by Eq.~\eqref{splitting}. This is why the multimode OAM-space test in Fig.~\ref{fig:multimode} is part of the argument rather than a decorative extra: it checks that the target doublet remains a clean effective subsystem.

A third limitation concerns the distinction between classical and fully quantum openness. The Hamiltonian $H_{\rm eff}$ describes resonance poles, classical wave amplitudes, scattering resonances, or postselected no-jump dynamics. For a genuinely quantum OAM device, the density matrix obeys a Lindblad master equation, and Hamiltonian EPs of the conditional no-jump Hamiltonian need not coincide with Liouvillian EPs of the full generator~\cite{Minganti2019,Arkhipov2020,Abo2024}. For this reason, the present theory is most directly applicable to classical photonic amplitudes, semiclassical fields, and spectroscopy-level observables. A fully quantum implementation should add one further step: after locating the amplitude EP, reconstruct or calculate the Liouvillian spectrum and determine whether the same geometric compensation also moves a Liouvillian EP.

Within that scope, the outlook is broad. The OAM-space viewpoint suggests immediate extensions to higher-order EPs built from more than two coupled OAM channels, to nonlinear photonic gain in which $g$ and $\gamma_i$ become intensity-dependent, and to full-vector electromagnetic or finite-element calculations that extract the same parameters from a concrete cavity or waveguide geometry. The present article should therefore be read as a compact theoretical proposal with measurable consequences: it identifies a geometry-controlled route to OAM photonic EPs and provides both analytic and measurement-level diagnostics for testing it.

\section{Conclusions}\label{secXI}

We have proposed a photonic route to exceptional points controlled by helicoidal geometry.
Starting from a torsion-free helicoidal metric, we identified a twist-induced chiral splitting between opposite-OAM modes. Upon opening the system, this splitting becomes the geometric part of the real detuning of an effective non-Hermitian OAM doublet. The exceptional-point condition is then reached when the twist compensates the residual detuning while the linewidth contrast satisfies the non-Hermitian balance condition.
In this precise operational sense, openness supplies the defective structure, and geometry supplies the control parameter.

Beyond the isolated $2\times2$ model, we embedded the target doublet in a multimode OAM manifold, verified the collapse of phase rigidity, demonstrated linear tracking of the EP position with residual detuning, and showed both static and dynamical encircling.
We further validated the mechanism in a device-level paraxial model of reciprocal and asymmetric-coupling Wiersig-type photonic platforms, extracting EP parameters from the full multimode device Hamiltonian rather than imposing them in a reduced two-mode block.
We then connected the theory to experimentally accessible photonic observables through a coupled-mode input-output response and three-dimensional spectral visualizations.
The resulting framework targets OAM photonics not only at the level of spectral poles but also at measurable cavity or waveguide signatures.

The central prediction is falsifiable. A geometry-controlled EP should reveal a calibrated real OAM splitting slope, a linear displacement of $\Omega_{\rm EP}$ with $\Delta_0$, eigenvalue and eigenvector coalescence, and orientation-dependent mode conversion under encircling. When these signatures are observed together in an OAM photonic platform, they would establish that helicoidal geometry can be used as a genuine control parameter for non-Hermitian singularities and their branch topology.

\begin{acknowledgments}
The author acknowledges support from Conselho Nacional de Desenvolvimento Cient\'{i}fico e Tecnol\'{o}gico (CNPq) (grants 306308/2022-3), Funda\c c\~ao de Amparo \`a Pesquisa e ao Desenvolvimento Cient\'{i}fico e Tecnol\'{o}gico do Maranh\~ao (FAPEMA) (grants UNIVERSAL-06395/22), and Coordena\c c\~ao de Aperfei\c coamento de Pessoal de N\'{i}vel Superior (CAPES) - Brazil (Finance Code 001).
\end{acknowledgments}

\end{document}